\pdfoutput=1 

\documentclass[sigconf,natbib=true]{source/acmart}
\AtBeginDocument{%
  }

\setcopyright{acmlicensed}
\copyrightyear{2018}
\acmYear{2018}
\acmDOI{XXXXXXX.XXXXXXX}
\acmConference[Conference acronym 'XX]{Make sure to enter the correct
  conference title from your rights confirmation emai}{June 03--05,
  2018}{Woodstock, NY}
\acmISBN{978-1-4503-XXXX-X/18/06}
\begin{document}

\title{Ranking-based Fusion Algorithms for Extreme Multi-label Text Classification (XMTC)}

\author{Celso França}
\email{celsofranca@dcc.ufmg.br}
\orcid{0000-0002-0251-7172}
\affiliation{%
  \institution{Federal University of Minas Gerais}
  \city{Belo Horizonte}
  \country{Brazil}
}

\author{Gestefane Rabbi}
\email{gestefane@dcc.ufmg.br}
\orcid{0009-0006-4484-3773}
\affiliation{%
  \institution{Federal University of Minas Gerais}
  \city{Belo Horizonte}
  \country{Brazil}
}

\author{Thiago Salles}
\email{tsalles@dcc.ufmg.br}
\orcid{0000-0003-2165-1999}
\affiliation{%
  \institution{Federal University of Minas Gerais}
  \city{Belo Horizonte}
  \country{Brazil}
}

\author{Washington Cunha}
\email{washingtoncunha@dcc.ufmg.br}
\orcid{0000-0002-1988-8412}
\affiliation{%
  \institution{Federal University of Minas Gerais}
  \city{Belo Horizonte}
  \country{Brazil}
}

\author{Leonardo Rocha}
\email{lcrocha@ufsj.edu.br}
\orcid{0000-0002-4913-4902}
\affiliation{%
  \institution{Federal University of São João del-Rei}
  \city{São João del-Rei}
  \country{Brazil}
}

\author{Marcos André Gonçalves}
\email{mgoncalv@dcc.ufmg.br}
\orcid{0000-0002-2075-3363}
\affiliation{%
  \institution{Federal University of Minas Gerais}
  \city{Belo Horizonte}
  \country{Brazil}
}
\renewcommand{\shortauthors}{França et al.}

\begin{abstract}
  In the context of Extreme Multi-label Text Classification (XMTC), where labels are assigned to text instances from a large label space, the long-tail distribution of labels presents a significant challenge. Labels can be broadly categorized into frequent, high-coverage \textbf{head labels} and infrequent, low-coverage \textbf{tail labels}, complicating the task of balancing effectiveness across all labels. To address this, combining predictions from multiple retrieval methods, such as sparse retrievers (e.g., BM25) and dense retrievers (e.g., fine-tuned BERT), offers a promising solution. The fusion of \textit{sparse} and \textit{dense} retrievers is motivated by the complementary ranking characteristics of these methods. Sparse retrievers compute relevance scores based on high-dimensional, bag-of-words representations, while dense retrievers utilize approximate nearest neighbor (ANN) algorithms on dense text and label embeddings within a shared embedding space. Rank-based fusion algorithms leverage these differences by combining the precise matching capabilities of sparse retrievers with the semantic richness of dense retrievers, thereby producing a final ranking that improves the effectiveness across both head and tail labels.
\end{abstract}



\received{20 February 2007}
\received[revised]{12 March 2009}
\received[accepted]{5 June 2009}

\maketitle

\section{Introduction}\label{sc_introduction}

Rank-based fusion algorithms~\cite{Bassani_SIGIR_2023} aim to assemble rankings from multiple retrieval systems to produce a unified and optimized final ranking. These algorithms perform under the assumption that aggregating evidence from diverse sources can enhance retrieval effectiveness, emphasizing consensus and mitigating individual system weaknesses. Rank-based fusion is particularly effective when the strengths of different retrieval methods are complementary.

In the context of Extreme Multi-label Text Classification (XMTC), which involves assigning labels to text instances from an extremely large label space, the long-tail distribution of labels poses a significant challenge. Specifically, labels can be broadly categorized into frequent, high-coverage head labels and infrequent, low-coverage tail labels, making it challenging to balance effectiveness across all labels. Therefore, combining predictions from multiple retrieval methods, such as sparse retrievers (e.g., BM25) and dense retrievers (e.g., fine-tuned BERT), has proven effective. The fusion process enables improved label ranking by leveraging the complementary strengths of sparse and dense methods, ensuring better coverage of both head and tail labels while maintaining overall ranking quality.

\section{Normalization Strategies}\label{sec_normalization_strategies}

The first step in fusing the results from multiple retrieval systems using score-based fusion methods is score normalization. This step ensures that relevance scores, which can vary significantly across retrieval models, are comparable. 

Normalization is crucial because different retrieval models compute relevance scores using diverse scales, intervals, and distributions. For instance, the classical probabilistic retrieval model BM25~\cite{Askari_2023} generates unbounded positive relevance scores. In contrast, modern deep learning-based retrieval systems, which rely on dot products or cosine similarity to compute relevance, often produce scores that are either unbounded or constrained within a fixed interval, such as $[-1, 1]$. To address these discrepancies, we leverage six normalization strategies: Min-Max Norm, Max Norm, Sum Norm, ZMUV Norm, Rank Norm, and Borda Norm~\cite{Renda_2003,Ranxfuse_2022,Bassani_SIGIR_2023}.

\paragraph{\textbf{Min-Max Norm}} Min-Max Norm scales the ranking scores to the interval $[0, 1]$, where the minimum score is transformed to $0$ and the maximum score is scaled to $1$.

\paragraph{\textbf{Max Norm}} Max Norm set the maximum score in the ranking to $1$  with all other scores scalled proportionally.

\paragraph{\textbf{Sum Norm}} Sum Norm adjusts ranking scores to ensure they are proportional and sum to $1$. It achieves this by first shifting all scores such that the minimum score becomes $0$. Then, the total sum of the adjusted scores is calculated, and each score is divided by this sum. 

\paragraph{\textbf{ZMUV Norm}} ZMUV Norm (Zero-Mean, Unit-Variance) normalizes ranking scores by transforming them into a standard normal distribution with a mean of $0$ and a variance of $1$. This is achieved by first calculating the mean and standard deviation of the scores. Each score is then adjusted by subtracting the mean and dividing by the standard deviation.

\paragraph{\textbf{Rank Norm}} Rank Norm normalizes ranking scores by assigning values based on their position in the ranked list. The top-ranked result is assigned a score of 1, while the bottom-ranked result gets a score of $\frac{1}{N}$, where $N$ is the total number of results. 

\paragraph{\textbf{Borda Norm}} Borda Norm normalizes ranking scores by assigning decreasing points to results based on their position in the ranking. The top position receives $N-1$ points, the next $N-2$, and so on, down to 0 points for the last position. The scores are then normalized by dividing by $N-1$. 
\section{Fusion Algorithms}\label{sec_fusion_algorithms}

We categorized the fusion algorithms into three categories: score-based methods~\cite{DBLP:conf/trec/FoxS93}, rank-based methods~\cite{DBLP:journals/cmig/MouraoMM15} and voting-based methods~\cite{DBLP:conf/sigir/AslamM01,DBLP:conf/cikm/MontagueA02}.

\subsection{Score-Based Methods}

\paragraph{\textbf{CombMIN}}
CombMIN selects the minimum ranking score assigned to each document across different ranking sources. The primary rationale behind this method is to reduce the likelihood of non-relevant documents being ranked highly, as it prioritizes caution by relying on the least favorable assessment from the contributing systems.

\paragraph{\textbf{CombMAX}}
CombMAX identifies the maximum ranking score assigned to each document across all retrieval results. This method ensures that relevant documents, even if highly ranked by only one retrieval system, are not overlooked due to poor performance by other systems.

\paragraph{\textbf{CombMED}}
CombMED calculates the median ranking score for each document across all retrieval results. By using the median, this method provides a robust central tendency measure, mitigating the impact of extreme scores and ensuring a balanced evaluation of relevance.

\paragraph{\textbf{CombSUM}}
CombSUM aggregates the ranking scores for each document by summing them across all retrieval systems. This straightforward approach assumes that higher cumulative scores correlate with higher relevance, thereby leveraging the collective contributions of multiple retrieval systems.

\paragraph{\textbf{CombANZ}}
CombANZ refines CombSUM by normalizing the summed ranking scores with the number of non-zero scores for each document. This normalization ensures fairness by accounting for differences in the number of retrieval systems contributing to the document's score.

\paragraph{\textbf{CombMNZ}}
CombMNZ builds upon CombSUM by multiplying the summed ranking score of each document by the count of non-zero scores. This method amplifies the influence of documents supported by multiple retrieval systems, highlighting consensus among diverse retrieval strategies.

\subsection{Rank-Based Methods}

\paragraph{\textbf{ISR}}
Information Synthesis Ranking (ISR) is a rank-based fusion method designed to aggregate evidence from multiple retrieval systems by assigning weights to documents proportional to their frequency of retrieval. The central assumption of ISR is that a document's relevance is positively correlated with the number of systems that identify it as relevant. By emphasizing consensus among systems, ISR aims to improve retrieval effectiveness by leveraging the collective agreement of diverse retrieval methods.

\paragraph{\textbf{Log-ISR}}
Log-Information Synthesis Ranking (Log-ISR) extends ISR by incorporating a logarithmic transformation into the weighting process. Instead of assigning weights directly proportional to retrieval frequency, Log-ISR scales these weights logarithmically. This adjustment mitigates the dominance of documents retrieved with exceptionally high frequency while preserving the relative importance of consensus across systems. The logarithmic scaling promotes a more balanced ranking, preventing the overrepresentation of outliers while maintaining the robustness of the fusion process.

\subsection{Voting-Based Methods}

\paragraph{\textbf{BordaFuse}}
\textbf{BordaFuse} is a rank-based fusion method derived from the Borda count, a voting system commonly employed in decision-making and preference aggregation. In BordaFuse, each document is assigned a cumulative score based on its rank across multiple retrieval systems. Specifically, a document's position in each ranked list contributes a score inversely proportional to its rank (e.g., higher ranks receive higher scores). The final score is obtained by summing these contributions across all retrieval systems, with higher scores indicating greater consensus. This approach effectively integrates the rankings from diverse retrieval methods to produce a comprehensive and robust final ranking.

\paragraph{\textbf{Condorcet}}
The \textbf{Condorcet} fusion method draws inspiration from the Condorcet voting principle, a cornerstone of social choice theory. It compares pairs of documents to determine which document is preferred over the other based on their relative rankings across multiple retrieval systems. Each pairwise comparison results in a "win" for the document ranked higher in the majority of systems. The final ranking is derived by aggregating these pairwise preferences, prioritizing documents with the highest overall wins. This method emphasizes relative ranking consensus, ensuring that documents consistently favored across systems are ranked higher in the final list.
\section{Experimental Setup}
\label{sec_experimental_setup}

To evaluate the effectiveness of the rank-based algorithms under the perspective of the three XMTC's challenges, we conducted extensive experiments leveraging the six normalization strategies outlined in Section~\ref{sec_normalization_strategies} and the ten rank-based fusion algorithms detailed in Section~\ref{sec_fusion_algorithms}.

These experiments were performed on four widely adopted and representative XMTC benchmarking datasets~\cite{HuiYe_2024, Jain_2023, Zhang_2021, Ronghui_2019, Jiang_2021}: \textbf{Eurlex-4K}, \textbf{Wiki10-31K}, \textbf{Amazon-670K}, and \textbf{AmazonCat-13K}. Table~\ref{tb_dataset_stats} summarizes the key statistics of these datasets, emphasizing their diversity in scale and structure. For example, \textbf{AmazonCat-13K} contains millions of documents, while \textbf{Amazon-670K} includes hundreds of thousands of labels. This diversity enables a robust and comprehensive evaluation of our approach across various scenarios.\looseness=-1

\begin{table}[ht]
\caption{Dataset statistics. \# of text instances (N);  \# of labels ($\bar{L}$); Average \# of relevant tail $(\bar{t})$ and  head $(\bar{h})$ labels; and Average \# of  instances per label($\bar{n}$).}
\label{tb_dataset_stats}
\centering
\begin{tabular}{r|rrrrr}
\hline
\textbf{Dataset}       & \multicolumn{1}{c}{N} & \multicolumn{1}{c}{$\bar{L}$} & \multicolumn{1}{c}{$\bar{t}$} & \multicolumn{1}{c}{$\bar{h}$} & \multicolumn{1}{c}{$\bar{n}$} \\ \hline
\textbf{Eurlex-4k}     & 19,314                         & 3956                           & 1.07                           & 4.25                           & 20.79                          \\
\textbf{Wiki10-31k}    & 20,762                         & 30,938                         & 3.66                           & 15.1                           & 8.52                           \\
\textbf{Amazon-670k}   & 643,474                        & 670,091                        & 2.56                           & 2.83                           & 3.99                           \\
\textbf{AmazonCat-13k} & 1,493,021                      & 13,330                         & 0.3                            & 4.75                           & 448.57                         \\ \hline
\end{tabular}

\end{table}

A formal definition is adopted to categorize the labels into tail and head. Let $\mathcal{T} = \{t_i , y_i\}^N_{i=1}$ be an XMTC dataset where the $N$ labels follow a long-tail distribution. Suppose labels $\{l_1, ..., l_L\}$ are organized by frequencies in descending order. By setting a threshold $h$, frequently occurring labels $\{l_1, ..., l_h\}$ are referred to as head labels, while infrequent ones $\{l_{h+1}, ..., l_L\}$ as tail labels. The Pareto principle determines the threshold $h$, categorizing the $80\%$ least frequent labels as tail and the remaining $20\%$ as head.\looseness=-1

For the sparse retriever, we set the main BM25 parameters at the retrieving stage as b=$0.75$ and k=$1.5$~\cite{Askari_2023}. Regarding the dense retriever, we fine-tune BERT to represent text and labels as embeddings into a shared vector space using the Normalized Temperature-Scaled Cross Entropy Loss~\cite{Jiajing_2023} as a learning objective. We let the training occur until there is no learning, which often occurs by the third epoch. We set the learning rate between $5e^{-3}$ and $5e^{-5}$ through a cyclical learning rate policy~\cite{WeiH_2022}. For sparse and dense retrievers, we set the number of candidates to be $64$ tail labels and $64$ head labels for both retrievers, producing $128$ candidates for the fusing stage. \looseness=-1
 
To ensure our results' robustness and reproducibility, we adopt a 5-fold cross-validation approach across all datasets. This rigorous experimental setup guarantees the generality of our observations and serves as a benchmark standard, enabling fair comparisons across different methods~\cite{Cunha_2023,pasin2024quantum}. It also mitigates the risk of dataset-specific biases inherent in single train-test splits~\cite{Cunha_TOIS_2024,cunha2021cost,cunha2025thorough}.\looseness=-1

Effectiveness is evaluated using the  \textit{precision@k} and \textit{nDCG@k}, averaged across the five test splits. Since the number of relevant labels (tail + head) per instance varies across datasets--from $6$ for Amazon-670K and AmazonCat-13K, to $7$ for Eurlex-4k, and up to $19$ for Wiki10-31k--the average number of relevant labels is approximately $10$. Therefore, evaluating ranking for $k \in \{1,5,10\}$ better represents the evaluation benchmark. \looseness=-1

Finally, we assess the statistical significance of our results by employing the two-sided paired Student's t-test with $95\%$ confidence to compare the averaged results~\cite{Cunha_2023}. In addition, we make our source code, developed models, and data publicly available to ensure the reproducibility of our experiments and results. \looseness=-1
\section{Experimental Results}\label{sc_experimental_results}

Tables~\ref{tb_norm_fuse_eurlex4k}, \ref{tb_norm_fuse_wiki10_31k}, \ref{tb_norm_fuse_amazon_670k}, and \ref{tb_norm_fuse_amazoncat_13k} present the effectiveness of combining six normalization strategies with ten rank-based fusion algorithm baselines, evaluated in terms of nDCG@k and Precision@k across the four XMTC benchmark datasets. These results provide comprehensive insights into the performance of each combination on the \textbf{Eurlex-4K}, \textbf{Wiki10-31K}, \textbf{Amazon-670K}, and \textbf{AmazonCat-13K} datasets.

\begin{table*}
\centering
\caption{nDCG@\textit{k} and Precision@\textit{k} of Normalization Strategies and Fusion Algorithms on \textbf{Eurlex-4k}.}
\label{tb_norm_fuse_eurlex4k}

\resizebox{\textwidth}{!}{

\begin{tabular}{l l | c c c | c c c | c c c | c c c }
\hline

 &  & \multicolumn{6}{c|}{\textbf{nDCG x 100}} & \multicolumn{6}{c}{\textbf{Precision x 100}} \\
 &  & \multicolumn{3}{c|}{\textbf{Tail label}} & \multicolumn{3}{c|}{\textbf{Head label}} & \multicolumn{3}{c|}{\textbf{Tail label}} & \multicolumn{3}{c}{\textbf{Head label}} \\
\textbf{Normalization } & \textbf{Algorithms} & \textbf{@1} & \textbf{@5} & \textbf{@10} & \textbf{@1} & \textbf{@5} & \textbf{@10} & \textbf{@1} & \textbf{@5} & \textbf{@10} & \textbf{@1} & \textbf{@5} & \textbf{@10} \\
\hline
\textbf{Min-Max Norm} & \textbf{CombMNZ} & \textbf{50.1(1.3)} & 27.3(0.5) & 27.0(0.4) & 80.5(1.1) & 62.8(0.5) & 65.4(0.5) & 50.1(1.3) & 20.0(0.4) & 11.6(0.2) & 80.5(1.1) & 53.5(0.3) & 33.6(0.2) \\
\textbf{Min-Max Norm} & \textbf{CombSUM} & \textbf{50.3(1.2)} & 27.8(0.5) & 27.4(0.4) & 80.5(1.1) & 62.7(0.4) & 65.6(0.5) & 50.3(1.2) & 20.5(0.4) & 11.8(0.2) & 80.5(1.1) & 53.5(0.3) & 33.7(0.3) \\
\textbf{Min-Max Norm} & \textbf{CombMIN} & 32.8(1.6) & 19.4(0.7) & 20.1(0.6) & 64.7(1.6) & 48.8(1.2) & 53.4(1.2) & 32.8(1.6) & 14.9(0.6) & 9.4(0.3) & 64.7(1.6) & 41.5(1.0) & 28.5(0.6) \\
\textbf{Min-Max Norm} & \textbf{CombMAX} & 37.5(0.2) & 24.6(0.4) & 24.7(0.3) & 56.9(1.1) & 52.8(0.5) & 58.2(0.6) & 37.5(0.2) & 19.4(0.4) & 11.5(0.2) & 56.9(1.1) & 48.0(0.5) & 32.6(0.3) \\
\textbf{Min-Max Norm} & \textbf{CombMED} & 35.2(1.6) & 22.6(0.7) & 23.1(0.5) & 66.3(1.6) & 54.6(1.1) & 59.2(0.9) & 35.2(1.6) & 17.9(0.5) & 11.0(0.2) & 66.3(1.6) & 47.8(0.8) & 32.0(0.4) \\
\textbf{Min-Max Norm} & \textbf{CombANZ} & 35.2(1.6) & 22.6(0.7) & 23.1(0.5) & 66.3(1.6) & 54.6(1.1) & 59.2(0.9) & 35.2(1.6) & 17.9(0.5) & 11.0(0.2) & 66.3(1.6) & 47.8(0.8) & 32.0(0.4) \\
\textbf{Min-Max Norm} & \textbf{ISR} & 46.9(0.9) & 26.7(0.4) & 26.5(0.4) & 76.0(1.1) & 60.6(0.6) & 64.3(0.6) & 46.9(0.9) & 20.2(0.4) & 11.7(0.2) & 76.0(1.1) & 52.5(0.4) & 34.0(0.2) \\
\textbf{Min-Max Norm} & \textbf{Log\_ISR} & 47.7(1.1) & 26.1(0.4) & 25.3(0.3) & 76.4(1.1) & 60.7(0.5) & 63.7(0.6) & 47.7(1.1) & 19.2(0.4) & 10.6(0.2) & 76.4(1.1) & 52.5(0.2) & 33.2(0.2) \\
\textbf{Min-Max Norm} & \textbf{BordaFuse} & 47.2(1.4) & 26.1(0.6) & 26.1(0.5) & 77.3(0.8) & 61.9(0.6) & 64.6(0.6) & 47.2(1.4) & 19.2(0.4) & 11.4(0.2) & 77.3(0.8) & 53.2(0.5) & 33.4(0.3) \\
\textbf{Min-Max Norm} & \textbf{Condorcet} & 38.3(1.0) & 21.4(0.5) & 21.5(0.5) & 69.7(1.2) & 54.4(0.7) & 57.3(0.8) & 38.3(1.0) & 15.9(0.4) & 9.5(0.3) & 69.7(1.2) & 46.3(0.6) & 29.7(0.4) \\
\hline
\textbf{ZMUV Norm} & \textbf{CombMNZ} & \textbf{51.5(1.7)} & \textbf{28.5(0.6)} & \textbf{28.0(0.6)} & \textbf{81.8(1.0)} & \textbf{64.6(0.7)} & \textbf{67.1(0.6)} & \textbf{51.5(1.7)} & \textbf{21.0(0.4)} & \textbf{12.1(0.3)} & \textbf{81.8(1.0)} & \textbf{55.2(0.5)} & \textbf{34.5(0.3)} \\
\textbf{ZMUV Norm} & \textbf{CombSUM} & \textbf{51.8(1.6)} & \textbf{28.2(0.7)} & \textbf{27.7(0.6)} & \textbf{82.3(0.9)} & \textbf{64.6(0.7)} & \textbf{66.7(0.8)} & \textbf{51.8(1.6)} & \textbf{20.6(0.5)} & \textbf{11.8(0.3)} & \textbf{82.3(0.9)} & \textbf{55.1(0.6)} & \textbf{34.0(0.4)} \\
\textbf{ZMUV Norm} & \textbf{CombMIN} & 27.1(2.2) & 18.0(0.8) & 18.8(0.6) & 46.9(1.4) & 45.9(1.1) & 49.7(1.0) & 27.1(2.2) & 14.5(0.5) & 9.2(0.2) & 46.9(1.4) & 42.0(0.9) & 27.8(0.5) \\
\textbf{ZMUV Norm} & \textbf{CombMAX} & 45.0(2.3) & 26.0(0.8) & 26.1(0.7) & 79.4(1.5) & 61.2(1.2) & 64.6(1.0) & 45.0(2.3) & 19.7(0.7) & 11.7(0.3) & 79.4(1.5) & 52.0(1.0) & 33.4(0.4) \\
\textbf{ZMUV Norm} & \textbf{CombMED} & 40.3(3.0) & 23.7(0.9) & 24.1(0.8) & 76.4(1.3) & 59.4(1.1) & 62.4(1.0) & 40.3(3.0) & 18.1(0.6) & 11.0(0.3) & 76.4(1.3) & 50.6(0.9) & 32.2(0.4) \\
\textbf{ZMUV Norm} & \textbf{CombANZ} & 40.3(3.0) & 23.7(0.9) & 24.1(0.8) & 76.4(1.3) & 59.4(1.1) & 62.4(1.0) & 40.3(3.0) & 18.1(0.6) & 11.0(0.3) & 76.4(1.3) & 50.6(0.9) & 32.2(0.4) \\
\textbf{ZMUV Norm} & \textbf{ISR} & 46.9(0.9) & 26.7(0.4) & 26.5(0.4) & 76.0(1.1) & 60.6(0.6) & 64.3(0.6) & 46.9(0.9) & 20.2(0.4) & 11.7(0.2) & 76.0(1.1) & 52.5(0.4) & 34.0(0.2) \\
\textbf{ZMUV Norm} & \textbf{Log\_ISR} & 47.7(1.1) & 26.1(0.4) & 25.3(0.3) & 76.4(1.1) & 60.7(0.5) & 63.7(0.6) & 47.7(1.1) & 19.2(0.4) & 10.6(0.2) & 76.4(1.1) & 52.5(0.2) & 33.2(0.2) \\
\textbf{ZMUV Norm} & \textbf{BordaFuse} & 47.2(1.4) & 26.1(0.6) & 26.1(0.5) & 77.3(0.8) & 61.9(0.6) & 64.6(0.6) & 47.2(1.4) & 19.2(0.4) & 11.4(0.2) & 77.3(0.8) & 53.2(0.5) & 33.4(0.3) \\
\textbf{ZMUV Norm} & \textbf{Condorcet} & 38.3(1.0) & 21.4(0.5) & 21.5(0.5) & 69.7(1.2) & 54.4(0.7) & 57.3(0.8) & 38.3(1.0) & 15.9(0.4) & 9.5(0.3) & 69.7(1.2) & 46.3(0.6) & 29.7(0.4) \\
\hline

\textbf{Max Norm} & \textbf{CombMNZ} & 50.0(1.1) & 26.9(0.4) & 26.3(0.3) & 81.1(1.0) & 63.2(0.5) & 65.4(0.5) & 50.0(1.1) & 19.5(0.3) & 11.1(0.2) & 81.1(1.0) & 53.8(0.3) & 33.4(0.2) \\
\textbf{Max Norm} & \textbf{CombSUM} & 50.0(1.1) & 26.9(0.4) & 26.3(0.3) & 81.1(1.0) & 63.2(0.5) & 65.4(0.5) & 50.0(1.1) & 19.5(0.3) & 11.1(0.2) & 81.1(1.0) & 53.8(0.3) & 33.4(0.2) \\
\textbf{Max Norm} & \textbf{CombMIN} & 32.1(1.4) & 18.8(0.7) & 19.2(0.6) & 64.4(1.8) & 46.3(1.2) & 50.3(1.1) & 32.1(1.4) & 14.3(0.6) & 8.9(0.3) & 64.4(1.8) & 38.5(0.9) & 26.3(0.4) \\
\textbf{Max Norm} & \textbf{CombMAX} & 37.8(0.5) & 24.5(0.4) & 24.5(0.4) & 57.0(0.9) & 52.8(0.5) & 57.8(0.5) & 37.8(0.5) & 19.3(0.4) & 11.3(0.3) & 57.0(0.9) & 48.0(0.5) & 32.3(0.3) \\
\textbf{Max Norm} & \textbf{CombMED} & 34.8(1.7) & 21.9(0.7) & 22.2(0.5) & 66.1(1.8) & 52.8(0.9) & 57.1(0.9) & 34.8(1.7) & 17.2(0.5) & 10.4(0.2) & 66.1(1.8) & 45.7(0.6) & 30.5(0.3) \\
\textbf{Max Norm} & \textbf{CombANZ} & 34.8(1.7) & 21.9(0.7) & 22.2(0.5) & 66.1(1.8) & 52.8(0.9) & 57.1(0.9) & 34.8(1.7) & 17.2(0.5) & 10.4(0.2) & 66.1(1.8) & 45.7(0.6) & 30.5(0.3) \\
\textbf{Max Norm} & \textbf{ISR} & 46.9(0.9) & 26.7(0.4) & 26.5(0.4) & 76.0(1.1) & 60.6(0.6) & 64.3(0.6) & 46.9(0.9) & 20.2(0.4) & 11.7(0.2) & 76.0(1.1) & 52.5(0.4) & 34.0(0.2) \\
\textbf{Max Norm} & \textbf{Log\_ISR} & 47.7(1.1) & 26.1(0.4) & 25.3(0.3) & 76.4(1.1) & 60.7(0.5) & 63.7(0.6) & 47.7(1.1) & 19.2(0.4) & 10.6(0.2) & 76.4(1.1) & 52.5(0.2) & 33.2(0.2) \\
\textbf{Max Norm} & \textbf{BordaFuse} & 47.2(1.4) & 26.1(0.6) & 26.1(0.5) & 77.3(0.8) & 61.9(0.6) & 64.6(0.6) & 47.2(1.4) & 19.2(0.4) & 11.4(0.2) & 77.3(0.8) & 53.2(0.5) & 33.4(0.3) \\
\textbf{Max Norm} & \textbf{Condorcet} & 38.3(1.0) & 21.4(0.5) & 21.5(0.5) & 69.7(1.2) & 54.4(0.7) & 57.3(0.8) & 38.3(1.0) & 15.9(0.4) & 9.5(0.3) & 69.7(1.2) & 46.3(0.6) & 29.7(0.4) \\
\hline

\textbf{Sum Norm} & \textbf{CombMNZ} & 48.8(1.0) & 26.6(0.4) & 26.0(0.3) & 80.3(1.2) & 62.4(0.6) & 65.0(0.5) & 48.8(1.0) & 19.4(0.3) & 11.1(0.2) & 80.3(1.2) & 53.2(0.3) & 33.4(0.2) \\
\textbf{Sum Norm} & \textbf{CombSUM} & 48.4(1.0) & 26.2(0.3) & 25.9(0.2) & 80.4(1.2) & 62.3(0.6) & 65.1(0.6) & 48.4(1.0) & 19.1(0.3) & 11.1(0.2) & 80.4(1.2) & 53.0(0.4) & 33.4(0.3) \\
\textbf{Sum Norm} & \textbf{CombMIN} & 16.3(1.0) & 13.9(0.6) & 15.8(0.5) & 56.5(2.1) & 47.1(1.3) & 52.2(1.4) & 16.3(1.0) & 12.2(0.5) & 8.8(0.3) & 56.5(2.1) & 41.2(1.1) & 28.7(0.6) \\
\textbf{Sum Norm} & \textbf{CombMAX} & 39.4(0.5) & 23.3(0.1) & 23.6(0.1) & 67.0(1.6) & 55.1(0.8) & 59.7(0.8) & 39.4(0.5) & 17.8(0.2) & 10.9(0.1) & 67.0(1.6) & 48.2(0.6) & 32.2(0.3) \\
\textbf{Sum Norm} & \textbf{CombMED} & 32.1(1.0) & 21.4(0.3) & 22.0(0.4) & 65.6(2.1) & 55.1(1.1) & 59.5(1.0) & 32.1(1.0) & 17.2(0.2) & 10.7(0.2) & 65.6(2.1) & 48.3(0.8) & 32.2(0.4) \\
\textbf{Sum Norm} & \textbf{CombANZ} & 32.1(1.0) & 21.4(0.3) & 22.0(0.4) & 65.6(2.1) & 55.1(1.1) & 59.5(1.0) & 32.1(1.0) & 17.2(0.2) & 10.7(0.2) & 65.6(2.1) & 48.3(0.8) & 32.2(0.4) \\
\textbf{Sum Norm} & \textbf{ISR} & 46.9(0.9) & 26.7(0.4) & 26.5(0.4) & 76.0(1.1) & 60.6(0.6) & 64.3(0.6) & 46.9(0.9) & 20.2(0.4) & 11.7(0.2) & 76.0(1.1) & 52.5(0.4) & 34.0(0.2) \\
\textbf{Sum Norm} & \textbf{Log\_ISR} & 47.7(1.1) & 26.1(0.4) & 25.3(0.3) & 76.4(1.1) & 60.7(0.5) & 63.7(0.6) & 47.7(1.1) & 19.2(0.4) & 10.6(0.2) & 76.4(1.1) & 52.5(0.2) & 33.2(0.2) \\
\textbf{Sum Norm} & \textbf{BordaFuse} & 47.2(1.4) & 26.1(0.6) & 26.1(0.5) & 77.3(0.8) & 61.9(0.6) & 64.6(0.6) & 47.2(1.4) & 19.2(0.4) & 11.4(0.2) & 77.3(0.8) & 53.2(0.5) & 33.4(0.3) \\
\textbf{Sum Norm} & \textbf{Condorcet} & 38.3(1.0) & 21.4(0.5) & 21.5(0.5) & 69.7(1.2) & 54.4(0.7) & 57.3(0.8) & 38.3(1.0) & 15.9(0.4) & 9.5(0.3) & 69.7(1.2) & 46.3(0.6) & 29.7(0.4) \\
\hline

\textbf{Rank Norm} & \textbf{CombMNZ} & 46.2(0.7) & 25.7(0.3) & 25.6(0.4) & 73.1(0.6) & 59.6(0.4) & 62.7(0.4) & 46.2(0.7) & 19.1(0.2) & 11.2(0.2) & 73.1(0.6) & 51.7(0.4) & 32.9(0.1) \\
\textbf{Rank Norm} & \textbf{CombSUM} & 46.3(0.7) & 25.9(0.3) & 26.1(0.4) & 73.1(0.6) & 59.6(0.5) & 62.9(0.4) & 46.3(0.7) & 19.3(0.2) & 11.6(0.2) & 73.1(0.6) & 51.7(0.4) & 33.1(0.2) \\
\textbf{Rank Norm} & \textbf{CombMIN} & 29.2(1.4) & 17.6(0.6) & 18.5(0.6) & 59.3(1.5) & 43.0(1.0) & 47.2(0.9) & 29.2(1.4) & 13.8(0.5) & 9.0(0.2) & 59.3(1.5) & 35.8(0.9) & 24.8(0.3) \\
\textbf{Rank Norm} & \textbf{CombMAX} & 36.7(0.5) & 24.5(0.5) & 24.8(0.5) & 58.6(0.9) & 57.2(0.7) & 60.7(0.8) & 36.7(0.5) & 19.2(0.6) & 11.6(0.3) & 58.6(0.9) & 51.2(0.7) & 32.9(0.5) \\
\textbf{Rank Norm} & \textbf{CombMED} & 31.4(1.4) & 19.8(0.7) & 20.6(0.6) & 63.3(1.2) & 49.8(1.0) & 53.7(1.0) & 31.4(1.4) & 15.6(0.5) & 9.9(0.2) & 63.3(1.2) & 42.6(0.8) & 28.5(0.4) \\
\textbf{Rank Norm} & \textbf{CombANZ} & 31.4(1.4) & 19.8(0.7) & 20.6(0.6) & 63.3(1.2) & 49.8(1.0) & 53.7(1.0) & 31.4(1.4) & 15.6(0.5) & 9.9(0.2) & 63.3(1.2) & 42.6(0.8) & 28.5(0.4) \\
\textbf{Rank Norm} & \textbf{ISR} & 46.9(0.9) & 26.7(0.4) & 26.5(0.4) & 76.0(1.1) & 60.6(0.6) & 64.3(0.6) & 46.9(0.9) & 20.2(0.4) & 11.7(0.2) & 76.0(1.1) & 52.5(0.4) & 34.0(0.2) \\
\textbf{Rank Norm} & \textbf{Log\_ISR} & 47.7(1.1) & 26.1(0.4) & 25.3(0.3) & 76.4(1.1) & 60.7(0.5) & 63.7(0.6) & 47.7(1.1) & 19.2(0.4) & 10.6(0.2) & 76.4(1.1) & 52.5(0.2) & 33.2(0.2) \\
\textbf{Rank Norm} & \textbf{BordaFuse} & 47.2(1.4) & 26.1(0.6) & 26.1(0.5) & 77.3(0.8) & 61.9(0.6) & 64.6(0.6) & 47.2(1.4) & 19.2(0.4) & 11.4(0.2) & 77.3(0.8) & 53.2(0.5) & 33.4(0.3) \\
\textbf{Rank Norm} & \textbf{Condorcet} & 38.3(1.0) & 21.4(0.5) & 21.5(0.5) & 69.7(1.2) & 54.4(0.7) & 57.3(0.8) & 38.3(1.0) & 15.9(0.4) & 9.5(0.3) & 69.7(1.2) & 46.3(0.6) & 29.7(0.4) \\
\hline

\textbf{Borda Norm} & \textbf{CombMNZ} & 47.0(1.6) & 26.1(0.6) & 26.1(0.5) & 78.0(1.1) & 62.0(0.7) & 64.7(0.6) & 47.0(1.6) & 19.2(0.4) & 11.4(0.2) & 78.0(1.1) & 53.2(0.5) & 33.4(0.3) \\
\textbf{Borda Norm} & \textbf{CombSUM} & 47.0(1.6) & 26.1(0.6) & 26.1(0.5) & 78.0(1.1) & 62.0(0.7) & 64.7(0.6) & 47.0(1.6) & 19.2(0.4) & 11.4(0.2) & 78.0(1.1) & 53.2(0.5) & 33.4(0.3) \\
\textbf{Borda Norm} & \textbf{CombMIN} & 45.9(2.1) & 25.1(0.6) & 24.6(0.6) & 77.0(0.5) & 60.6(0.6) & 63.3(0.6) & 45.9(2.1) & 18.4(0.4) & 10.4(0.2) & 77.0(0.5) & 51.7(0.5) & 32.7(0.3) \\
\textbf{Borda Norm} & \textbf{CombMAX} & 39.3(0.9) & 25.2(0.5) & 25.1(0.4) & 68.3(0.8) & 58.3(0.6) & 62.3(0.6) & 39.3(0.9) & 19.7(0.4) & 11.5(0.3) & 68.3(0.8) & 51.2(0.5) & 33.4(0.3) \\
\textbf{Borda Norm} & \textbf{CombMED} & 47.0(1.6) & 26.1(0.6) & 26.1(0.5) & 78.0(1.1) & 62.0(0.7) & 64.7(0.6) & 47.0(1.6) & 19.2(0.4) & 11.4(0.2) & 78.0(1.1) & 53.2(0.5) & 33.4(0.3) \\
\textbf{Borda Norm} & \textbf{CombANZ} & 47.0(1.6) & 26.1(0.6) & 26.1(0.5) & 78.0(1.1) & 62.0(0.7) & 64.7(0.6) & 47.0(1.6) & 19.2(0.4) & 11.4(0.2) & 78.0(1.1) & 53.2(0.5) & 33.4(0.3) \\
\textbf{Borda Norm} & \textbf{ISR} & 1.3(0.5) & 1.4(0.2) & 2.0(0.2) & 4.0(0.2) & 3.7(0.3) & 5.7(0.3) & 1.3(0.5) & 1.3(0.1) & 1.3(0.1) & 4.0(0.2) & 3.6(0.3) & 4.0(0.2) \\
\textbf{Borda Norm} & \textbf{Log\_ISR} & 1.3(0.5) & 1.4(0.2) & 2.0(0.2) & 4.0(0.2) & 3.7(0.3) & 5.7(0.3) & 1.3(0.5) & 1.3(0.1) & 1.3(0.1) & 4.0(0.2) & 3.6(0.3) & 4.0(0.2) \\
\textbf{Borda Norm} & \textbf{BordaFuse} & 1.3(0.5) & 1.4(0.2) & 2.0(0.2) & 4.0(0.2) & 3.7(0.3) & 5.7(0.3) & 1.3(0.5) & 1.3(0.1) & 1.3(0.1) & 4.0(0.2) & 3.6(0.3) & 4.0(0.2) \\
\textbf{Borda Norm} & \textbf{Condorcet} & 1.3(0.5) & 1.4(0.2) & 2.0(0.2) & 4.0(0.2) & 3.7(0.3) & 5.7(0.3) & 1.3(0.5) & 1.3(0.1) & 1.3(0.1) & 4.0(0.2) & 3.6(0.3) & 4.0(0.2) \\

\hline

\end{tabular}
}
\end{table*}

\begin{table*}
\centering
\caption{nDCG@\textit{k} and Precision@\textit{k} of Normalization Strategies and Fusion Algorithms on \textbf{Wiki10-31k}.}
\label{tb_norm_fuse_wiki10_31k}

\resizebox{\textwidth}{!}{

\begin{tabular}{l l | c c c | c c c | c c c | c c c}

\hline

 &  & \multicolumn{6}{c|}{\textbf{nDCG x 100}} & \multicolumn{6}{c}{\textbf{Precision x 100}} \\
 &  & \multicolumn{3}{c|}{\textbf{Tail label}} & \multicolumn{3}{c|}{\textbf{Head label}} & \multicolumn{3}{c|}{\textbf{Tail label}} & \multicolumn{3}{c}{\textbf{Head label}} \\

\textbf{Normalization } & \textbf{Algorithms} & \textbf{@1} & \textbf{@5} & \textbf{@10} & \textbf{@1} & \textbf{@5} & \textbf{@10} & \textbf{@1} & \textbf{@5} & \textbf{@10} & \textbf{@1} & \textbf{@5} & \textbf{@10} \\
\hline
\textbf{Min-Max Norm} & \textbf{CombMNZ} & \textbf{48.0(0.9)} & \textbf{26.5(0.4)} & \textbf{26.2(0.5)} & \textbf{80.2(1.4)} & \textbf{62.6(1.2)} & \textbf{65.2(1.0)} & \textbf{48.0(0.9)} & \textbf{19.5(0.3)} & \textbf{11.3(0.3)} & \textbf{80.2(1.4)} & \textbf{53.4(1.1)} & \textbf{33.5(0.5)} \\
\textbf{Min-Max Norm} & \textbf{CombSUM} & \textbf{48.2(1.1)} & \textbf{26.8(0.7)} & \textbf{26.4(0.7)} & \textbf{80.1(1.4)} & \textbf{62.6(1.1)} & \textbf{65.4(1.1)} & \textbf{48.2(1.1)} & \textbf{19.9(0.5)} & \textbf{11.4(0.3)} & \textbf{80.1(1.4)} & \textbf{53.4(1.1)} & \textbf{33.6(0.6)} \\
\textbf{Min-Max Norm} & \textbf{CombMIN} & 33.5(3.0) & 19.5(1.6) & 20.0(1.6) & 65.2(2.9) & 49.5(2.5) & 53.9(2.5) & 33.5(3.0) & 14.9(1.2) & 9.4(0.6) & 65.2(2.9) & 42.1(2.3) & 28.7(1.2) \\
\textbf{Min-Max Norm} & \textbf{CombMAX} & 37.6(0.7) & 24.0(0.8) & 24.1(0.7) & 57.0(1.2) & 52.8(0.8) & 58.0(0.9) & 37.6(0.7) & 18.9(0.7) & 11.2(0.4) & 57.0(1.2) & 48.0(0.8) & 32.5(0.5) \\
\textbf{Min-Max Norm} & \textbf{CombMED} & 35.6(3.2) & 22.3(1.4) & 22.8(1.3) & 66.8(2.7) & 55.1(1.9) & 59.5(1.8) & 35.6(3.2) & 17.5(0.9) & 10.8(0.4) & 66.8(2.7) & 48.2(1.7) & 32.0(0.7) \\
\textbf{Min-Max Norm} & \textbf{CombANZ} & 35.6(3.2) & 22.3(1.4) & 22.8(1.3) & 66.8(2.7) & 55.1(1.9) & 59.5(1.8) & 35.6(3.2) & 17.5(0.9) & 10.8(0.4) & 66.8(2.7) & 48.2(1.7) & 32.0(0.7) \\
\textbf{Min-Max Norm} & \textbf{ISR} & 45.7(2.2) & 25.9(1.0) & 25.7(1.0) & 75.7(2.1) & 60.3(1.7) & 64.1(1.5) & 45.7(2.2) & 19.6(0.7) & 11.4(0.4) & 75.7(2.1) & 52.3(1.6) & 33.9(0.7) \\
\textbf{Min-Max Norm} & \textbf{Log\_ISR} & 46.0(1.3) & 25.5(0.5) & 24.8(0.6) & 76.1(2.2) & 60.5(1.4) & 63.5(1.2) & 46.0(1.3) & 18.8(0.4) & 10.5(0.2) & 76.1(2.2) & 52.2(1.3) & 33.2(0.5) \\
\textbf{Min-Max Norm} & \textbf{BordaFuse} & 45.4(1.2) & 25.3(0.8) & 25.3(0.8) & 76.8(1.5) & 61.6(1.4) & 64.3(1.3) & 45.4(1.2) & 18.8(0.6) & 11.1(0.4) & 76.8(1.5) & 53.0(1.3) & 33.4(0.7) \\
\textbf{Min-Max Norm} & \textbf{Condorcet} & 37.9(2.1) & 21.2(1.1) & 21.3(1.0) & 69.4(2.6) & 54.5(1.8) & 57.4(1.8) & 37.9(2.1) & 15.8(0.8) & 9.4(0.4) & 69.4(2.6) & 46.6(1.6) & 29.8(0.9) \\
\hline

\textbf{ZMUV Norm} & \textbf{CombMNZ} & \textbf{49.1(1.6)} & \textbf{27.4(0.9)} & \textbf{27.0(0.9)} & \textbf{81.2(1.5)} & \textbf{64.2(1.6)} & \textbf{66.9(1.4)} & \textbf{49.1(1.6)} & \textbf{20.3(0.8)} & \textbf{11.7(0.4)} & \textbf{81.2(1.5)} & \textbf{54.9(1.5)} & \textbf{34.4(0.6)} \\
\textbf{ZMUV Norm} & \textbf{CombSUM} & \textbf{49.5(2.9)} & \textbf{27.2(1.4)} & \textbf{26.8(1.3)} & \textbf{81.8(1.7)} & \textbf{64.3(1.9)} & \textbf{66.6(1.7)} & \textbf{49.5(2.9)} & \textbf{20.0(1.1)} & \textbf{11.5(0.5)} & \textbf{81.8(1.7)} & \textbf{54.9(1.8)} & \textbf{34.0(0.8)} \\
\textbf{ZMUV Norm} & \textbf{CombMIN} & 27.9(3.9) & 18.3(1.8) & 18.9(1.7) & 49.0(4.8) & 47.2(3.4) & 50.6(2.9) & 27.9(3.9) & 14.7(1.2) & 9.2(0.6) & 49.0(4.8) & 42.9(2.8) & 28.0(1.1) \\
\textbf{ZMUV Norm} & \textbf{CombMAX} & 43.6(4.2) & 25.3(1.8) & 25.3(1.7) & 79.0(3.0) & 61.3(2.9) & 64.6(2.3) & 43.6(4.2) & 19.2(1.1) & 11.4(0.5) & 79.0(3.0) & 52.2(2.7) & 33.4(0.9) \\
\textbf{ZMUV Norm} & \textbf{CombMED} & 40.2(4.5) & 23.4(1.9) & 23.8(1.8) & 76.6(3.4) & 59.9(3.0) & 62.6(2.5) & 40.2(4.5) & 17.8(1.2) & 10.9(0.5) & 76.6(3.4) & 51.0(2.7) & 32.3(1.1) \\
\textbf{ZMUV Norm} & \textbf{CombANZ} & 40.2(4.5) & 23.4(1.9) & 23.8(1.8) & 76.6(3.4) & 59.9(3.0) & 62.6(2.5) & 40.2(4.5) & 17.8(1.2) & 10.9(0.5) & 76.6(3.4) & 51.0(2.7) & 32.3(1.1) \\
\textbf{ZMUV Norm} & \textbf{ISR} & 45.7(2.2) & 25.9(1.0) & 25.7(1.0) & 75.7(2.1) & 60.3(1.7) & 64.1(1.5) & 45.7(2.2) & 19.6(0.7) & 11.4(0.4) & 75.7(2.1) & 52.3(1.6) & 33.9(0.7) \\
\textbf{ZMUV Norm} & \textbf{Log\_ISR} & 46.0(1.3) & 25.5(0.5) & 24.8(0.6) & 76.1(2.2) & 60.5(1.4) & 63.5(1.2) & 46.0(1.3) & 18.8(0.4) & 10.5(0.2) & 76.1(2.2) & 52.2(1.3) & 33.2(0.5) \\
\textbf{ZMUV Norm} & \textbf{BordaFuse} & 45.4(1.2) & 25.3(0.8) & 25.3(0.8) & 76.8(1.5) & 61.6(1.4) & 64.3(1.3) & 45.4(1.2) & 18.8(0.6) & 11.1(0.4) & 76.8(1.5) & 53.0(1.3) & 33.4(0.7) \\
\textbf{ZMUV Norm} & \textbf{Condorcet} & 37.9(2.1) & 21.2(1.1) & 21.3(1.0) & 69.4(2.6) & 54.5(1.8) & 57.4(1.8) & 37.9(2.1) & 15.8(0.8) & 9.4(0.4) & 69.4(2.6) & 46.6(1.6) & 29.8(0.9) \\
\hline

\textbf{Max Norm} & \textbf{CombMNZ} & 48.1(1.5) & 26.2(0.4) & 25.6(0.5) & 80.6(1.2) & 63.0(1.1) & 65.2(0.9) & 48.1(1.5) & 19.2(0.3) & 10.9(0.2) & 80.6(1.2) & 53.6(1.0) & 33.3(0.4) \\
\textbf{Max Norm} & \textbf{CombSUM} & 48.1(1.5) & 26.2(0.4) & 25.6(0.5) & 80.6(1.2) & 63.0(1.1) & 65.2(0.9) & 48.1(1.5) & 19.2(0.3) & 10.9(0.2) & 80.6(1.2) & 53.6(1.0) & 33.3(0.4) \\
\textbf{Max Norm} & \textbf{CombMIN} & 32.8(3.2) & 18.9(1.6) & 19.2(1.5) & 64.8(3.2) & 46.9(3.0) & 50.7(3.0) & 32.8(3.2) & 14.4(1.2) & 8.8(0.6) & 64.8(3.2) & 39.1(2.8) & 26.4(1.6) \\
\textbf{Max Norm} & \textbf{CombMAX} & 37.4(1.0) & 24.0(0.9) & 24.0(0.9) & 56.9(1.2) & 52.7(0.9) & 57.6(1.0) & 37.4(1.0) & 18.7(0.7) & 11.0(0.4) & 56.9(1.2) & 47.9(0.8) & 32.2(0.6) \\
\textbf{Max Norm} & \textbf{CombMED} & 35.2(3.4) & 21.7(1.4) & 22.0(1.3) & 66.6(2.7) & 53.2(2.3) & 57.3(2.0) & 35.2(3.4) & 16.9(0.8) & 10.2(0.3) & 66.6(2.7) & 46.0(2.0) & 30.5(0.9) \\
\textbf{Max Norm} & \textbf{CombANZ} & 35.2(3.4) & 21.7(1.4) & 22.0(1.3) & 66.6(2.7) & 53.2(2.3) & 57.3(2.0) & 35.2(3.4) & 16.9(0.8) & 10.2(0.3) & 66.6(2.7) & 46.0(2.0) & 30.5(0.9) \\
\textbf{Max Norm} & \textbf{ISR} & 45.7(2.2) & 25.9(1.0) & 25.7(1.0) & 75.7(2.1) & 60.3(1.7) & 64.1(1.5) & 45.7(2.2) & 19.6(0.7) & 11.4(0.4) & 75.7(2.1) & 52.3(1.6) & 33.9(0.7) \\
\textbf{Max Norm} & \textbf{Log\_ISR} & 46.0(1.3) & 25.5(0.5) & 24.8(0.6) & 76.1(2.2) & 60.5(1.4) & 63.5(1.2) & 46.0(1.3) & 18.8(0.4) & 10.5(0.2) & 76.1(2.2) & 52.2(1.3) & 33.2(0.5) \\
\textbf{Max Norm} & \textbf{BordaFuse} & 45.4(1.2) & 25.3(0.8) & 25.3(0.8) & 76.8(1.5) & 61.6(1.4) & 64.3(1.3) & 45.4(1.2) & 18.8(0.6) & 11.1(0.4) & 76.8(1.5) & 53.0(1.3) & 33.4(0.7) \\
\textbf{Max Norm} & \textbf{Condorcet} & 37.9(2.1) & 21.2(1.1) & 21.3(1.0) & 69.4(2.6) & 54.5(1.8) & 57.4(1.8) & 37.9(2.1) & 15.8(0.8) & 9.4(0.4) & 69.4(2.6) & 46.6(1.6) & 29.8(0.9) \\
\hline

\textbf{Sum Norm} & \textbf{CombMNZ} & 47.5(1.2) & 26.0(0.4) & 25.5(0.4) & 80.1(1.3) & 62.3(1.1) & 64.9(1.0) & 47.5(1.2) & 19.1(0.2) & 10.9(0.2) & 80.1(1.3) & 53.0(1.1) & 33.3(0.5) \\
\textbf{Sum Norm} & \textbf{CombSUM} & 47.2(1.2) & 25.7(0.4) & 25.4(0.5) & 80.1(1.3) & 62.2(1.2) & 65.0(1.1) & 47.2(1.2) & 18.9(0.3) & 10.9(0.2) & 80.1(1.3) & 52.9(1.1) & 33.4(0.6) \\
\textbf{Sum Norm} & \textbf{CombMIN} & 18.3(2.4) & 14.9(1.3) & 16.5(1.3) & 58.5(3.5) & 48.5(2.8) & 53.3(2.8) & 18.3(2.4) & 13.0(1.0) & 8.9(0.6) & 58.5(3.5) & 42.3(2.4) & 29.0(1.4) \\
\textbf{Sum Norm} & \textbf{CombMAX} & 39.0(0.7) & 23.0(0.2) & 23.3(0.2) & 67.3(1.8) & 55.3(1.0) & 59.8(1.1) & 39.0(0.7) & 17.6(0.2) & 10.7(0.2) & 67.3(1.8) & 48.3(0.9) & 32.3(0.5) \\
\textbf{Sum Norm} & \textbf{CombMED} & 33.0(1.8) & 21.5(0.6) & 22.0(0.7) & 66.5(3.3) & 55.7(1.9) & 59.9(1.8) & 33.0(1.8) & 17.1(0.3) & 10.5(0.3) & 66.5(3.3) & 48.7(1.5) & 32.2(0.7) \\
\textbf{Sum Norm} & \textbf{CombANZ} & 33.0(1.8) & 21.5(0.6) & 22.0(0.7) & 66.5(3.3) & 55.7(1.9) & 59.9(1.8) & 33.0(1.8) & 17.1(0.3) & 10.5(0.3) & 66.5(3.3) & 48.7(1.5) & 32.2(0.7) \\
\textbf{Sum Norm} & \textbf{ISR} & 45.7(2.2) & 25.9(1.0) & 25.7(1.0) & 75.7(2.1) & 60.3(1.7) & 64.1(1.5) & 45.7(2.2) & 19.6(0.7) & 11.4(0.4) & 75.7(2.1) & 52.3(1.6) & 33.9(0.7) \\
\textbf{Sum Norm} & \textbf{Log\_ISR} & 46.0(1.3) & 25.5(0.5) & 24.8(0.6) & 76.1(2.2) & 60.5(1.4) & 63.5(1.2) & 46.0(1.3) & 18.8(0.4) & 10.5(0.2) & 76.1(2.2) & 52.2(1.3) & 33.2(0.5) \\
\textbf{Sum Norm} & \textbf{BordaFuse} & 45.4(1.2) & 25.3(0.8) & 25.3(0.8) & 76.8(1.5) & 61.6(1.4) & 64.3(1.3) & 45.4(1.2) & 18.8(0.6) & 11.1(0.4) & 76.8(1.5) & 53.0(1.3) & 33.4(0.7) \\
\textbf{Sum Norm} & \textbf{Condorcet} & 37.9(2.1) & 21.2(1.1) & 21.3(1.0) & 69.4(2.6) & 54.5(1.8) & 57.4(1.8) & 37.9(2.1) & 15.8(0.8) & 9.4(0.4) & 69.4(2.6) & 46.6(1.6) & 29.8(0.9) \\
\hline

\textbf{Rank Norm} & \textbf{CombMNZ} & 44.4(0.6) & 25.1(0.4) & 24.9(0.5) & 72.4(1.6) & 59.3(0.9) & 62.5(0.9) & 44.4(0.6) & 18.7(0.4) & 10.9(0.2) & 72.4(1.6) & 51.5(0.9) & 32.9(0.4) \\
\textbf{Rank Norm} & \textbf{CombSUM} & 44.4(0.6) & 25.2(0.5) & 25.3(0.6) & 72.4(1.6) & 59.3(0.9) & 62.6(0.9) & 44.4(0.6) & 18.8(0.5) & 11.2(0.3) & 72.4(1.6) & 51.5(0.9) & 33.0(0.5) \\
\textbf{Rank Norm} & \textbf{CombMIN} & 30.0(3.1) & 17.7(1.6) & 18.5(1.5) & 59.7(3.6) & 43.6(2.5) & 47.6(2.3) & 30.0(3.1) & 13.8(1.2) & 8.9(0.6) & 59.7(3.6) & 36.2(2.0) & 25.0(1.0) \\
\textbf{Rank Norm} & \textbf{CombMAX} & 36.9(0.5) & 24.0(1.2) & 24.3(1.1) & 58.5(1.4) & 57.3(2.4) & 60.7(2.0) & 36.9(0.5) & 18.7(1.2) & 11.3(0.6) & 58.5(1.4) & 51.4(2.6) & 32.9(1.1) \\
\textbf{Rank Norm} & \textbf{CombMED} & 32.1(3.0) & 19.9(1.6) & 20.5(1.5) & 63.2(3.2) & 50.3(2.4) & 54.0(2.2) & 32.1(3.0) & 15.6(1.2) & 9.8(0.5) & 63.2(3.2) & 43.1(2.2) & 28.6(1.0) \\
\textbf{Rank Norm} & \textbf{CombANZ} & 32.1(3.0) & 19.9(1.6) & 20.5(1.5) & 63.2(3.2) & 50.3(2.4) & 54.0(2.2) & 32.1(3.0) & 15.6(1.2) & 9.8(0.5) & 63.2(3.2) & 43.1(2.2) & 28.6(1.0) \\
\textbf{Rank Norm} & \textbf{ISR} & 45.7(2.2) & 25.9(1.0) & 25.7(1.0) & 75.7(2.1) & 60.3(1.7) & 64.1(1.5) & 45.7(2.2) & 19.6(0.7) & 11.4(0.4) & 75.7(2.1) & 52.3(1.6) & 33.9(0.7) \\
\textbf{Rank Norm} & \textbf{Log\_ISR} & 46.0(1.3) & 25.5(0.5) & 24.8(0.6) & 76.1(2.2) & 60.5(1.4) & 63.5(1.2) & 46.0(1.3) & 18.8(0.4) & 10.5(0.2) & 76.1(2.2) & 52.2(1.3) & 33.2(0.5) \\
\textbf{Rank Norm} & \textbf{BordaFuse} & 45.4(1.2) & 25.3(0.8) & 25.3(0.8) & 76.8(1.5) & 61.6(1.4) & 64.3(1.3) & 45.4(1.2) & 18.8(0.6) & 11.1(0.4) & 76.8(1.5) & 53.0(1.3) & 33.4(0.7) \\
\textbf{Rank Norm} & \textbf{Condorcet} & 37.9(2.1) & 21.2(1.1) & 21.3(1.0) & 69.4(2.6) & 54.5(1.8) & 57.4(1.8) & 37.9(2.1) & 15.8(0.8) & 9.4(0.4) & 69.4(2.6) & 46.6(1.6) & 29.8(0.9) \\
\hline

\textbf{Borda Norm} & \textbf{CombMNZ} & 45.5(1.3) & 25.3(0.7) & 25.3(0.9) & 77.6(1.6) & 61.7(1.4) & 64.5(1.4) & 45.5(1.3) & 18.8(0.6) & 11.1(0.4) & 77.6(1.6) & 52.9(1.4) & 33.4(0.7) \\
\textbf{Borda Norm} & \textbf{CombSUM} & 45.5(1.3) & 25.3(0.7) & 25.3(0.9) & 77.6(1.6) & 61.7(1.4) & 64.5(1.4) & 45.5(1.3) & 18.8(0.6) & 11.1(0.4) & 77.6(1.6) & 52.9(1.4) & 33.4(0.7) \\
\textbf{Borda Norm} & \textbf{CombMIN} & 44.3(1.3) & 24.5(0.7) & 24.1(0.7) & 76.6(1.3) & 60.3(1.4) & 63.2(1.4) & 44.3(1.3) & 18.1(0.6) & 10.3(0.3) & 76.6(1.3) & 51.5(1.4) & 32.7(0.7) \\
\textbf{Borda Norm} & \textbf{CombMAX} & 39.2(2.3) & 24.6(1.1) & 24.6(1.1) & 67.7(2.0) & 58.2(1.8) & 62.2(1.6) & 39.2(2.3) & 19.1(0.8) & 11.3(0.4) & 67.7(2.0) & 51.2(1.7) & 33.4(0.7) \\
\textbf{Borda Norm} & \textbf{CombMED} & 45.5(1.3) & 25.3(0.7) & 25.3(0.9) & 77.6(1.6) & 61.7(1.4) & 64.5(1.4) & 45.5(1.3) & 18.8(0.6) & 11.1(0.4) & 77.6(1.6) & 52.9(1.4) & 33.4(0.7) \\
\textbf{Borda Norm} & \textbf{CombANZ} & 45.5(1.3) & 25.3(0.7) & 25.3(0.9) & 77.6(1.6) & 61.7(1.4) & 64.5(1.4) & 45.5(1.3) & 18.8(0.6) & 11.1(0.4) & 77.6(1.6) & 52.9(1.4) & 33.4(0.7) \\
\textbf{Borda Norm} & \textbf{ISR} & 1.4(0.4) & 1.4(0.2) & 2.0(0.2) & 4.1(0.3) & 3.8(0.2) & 5.8(0.3) & 1.4(0.4) & 1.3(0.1) & 1.3(0.1) & 4.1(0.3) & 3.7(0.2) & 4.1(0.2) \\
\textbf{Borda Norm} & \textbf{Log\_ISR} & 1.4(0.4) & 1.4(0.2) & 2.0(0.2) & 4.1(0.3) & 3.8(0.2) & 5.8(0.3) & 1.4(0.4) & 1.3(0.1) & 1.3(0.1) & 4.1(0.3) & 3.7(0.2) & 4.1(0.2) \\
\textbf{Borda Norm} & \textbf{BordaFuse} & 1.4(0.4) & 1.4(0.2) & 2.0(0.2) & 4.1(0.3) & 3.8(0.2) & 5.8(0.3) & 1.4(0.4) & 1.3(0.1) & 1.3(0.1) & 4.1(0.3) & 3.7(0.2) & 4.1(0.2) \\
\textbf{Borda Norm} & \textbf{Condorcet} & 1.4(0.4) & 1.4(0.2) & 2.0(0.2) & 4.1(0.3) & 3.8(0.2) & 5.8(0.3) & 1.4(0.4) & 1.3(0.1) & 1.3(0.1) & 4.1(0.3) & 3.7(0.2) & 4.1(0.2) \\
\hline
\end{tabular}
}
\end{table*}

\begin{table*}
\centering
\caption{nDCG@\textit{k} and Precision@\textit{k} of Normalization Strategies and Fusion Algorithms on \textbf{Amazon-670k}.}
\label{tb_norm_fuse_amazon_670k}

\resizebox{\textwidth}{!}{

\begin{tabular}{l l | c c c | c c c | c c c | c c c}

\hline

 &  & \multicolumn{6}{c|}{\textbf{nDCG x 100}} & \multicolumn{6}{c}{\textbf{Precision x 100}} \\
 &  & \multicolumn{3}{c|}{\textbf{Tail label}} & \multicolumn{3}{c|}{\textbf{Head label}} & \multicolumn{3}{c|}{\textbf{Tail label}} & \multicolumn{3}{c}{\textbf{Head label}} \\

\textbf{Normalization } & \textbf{Algorithms} & \textbf{@1} & \textbf{@5} & \textbf{@10} & \textbf{@1} & \textbf{@5} & \textbf{@10} & \textbf{@1} & \textbf{@5} & \textbf{@10} & \textbf{@1} & \textbf{@5} & \textbf{@10} \\
\hline
\textbf{Min-Max Norm} & \textbf{CombMNZ} & 43.5(0.2) & 34.2(0.1) & 33.7(0.1) & 45.4(0.2) & 36.6(0.1) & 37.0(0.1) & 43.5(0.2) & 28.5(0.1) & 17.3(0.0) & 45.4(0.2) & 32.4(0.1) & 21.1(0.0) \\
\textbf{Min-Max Norm} & \textbf{CombSUM} & 43.4(0.1) & 33.7(0.1) & 33.2(0.1) & 45.3(0.2) & 36.5(0.1) & 36.8(0.1) & 43.4(0.1) & 27.8(0.1) & 16.9(0.1) & 45.3(0.2) & 32.2(0.1) & 20.9(0.0) \\
\textbf{Min-Max Norm} & \textbf{CombMIN} & 25.5(0.2) & 21.4(0.2) & 22.8(0.1) & 32.0(0.1) & 25.2(0.1) & 26.4(0.1) & 25.5(0.2) & 18.4(0.1) & 12.8(0.1) & 32.0(0.1) & 22.2(0.1) & 15.4(0.1) \\
\textbf{Min-Max Norm} & \textbf{CombMAX} & 28.4(0.1) & 27.2(0.1) & 28.2(0.1) & 36.3(0.1) & 32.8(0.1) & 33.8(0.1) & 28.4(0.1) & 24.2(0.1) & 16.0(0.1) & 36.3(0.1) & 30.1(0.1) & 20.0(0.1) \\
\textbf{Min-Max Norm} & \textbf{CombMED} & 26.0(0.2) & 23.9(0.1) & 25.6(0.1) & 33.9(0.2) & 29.6(0.1) & 31.3(0.1) & 26.0(0.2) & 21.3(0.1) & 14.9(0.1) & 33.9(0.2) & 27.0(0.1) & 18.9(0.1) \\
\textbf{Min-Max Norm} & \textbf{CombANZ} & 26.0(0.2) & 23.9(0.1) & 25.6(0.1) & 33.9(0.2) & 29.6(0.1) & 31.3(0.1) & 26.0(0.2) & 21.3(0.1) & 14.9(0.1) & 33.9(0.2) & 27.0(0.1) & 18.9(0.1) \\
\textbf{Min-Max Norm} & \textbf{ISR} & 43.7(0.2) & 32.7(0.1) & 32.9(0.1) & 43.4(0.1) & 34.2(0.1) & 35.5(0.1) & 43.7(0.2) & 27.2(0.1) & 17.2(0.1) & 43.4(0.1) & 30.4(0.1) & 20.7(0.0) \\
\textbf{Min-Max Norm} & \textbf{Log\_ISR} & 42.7(0.2) & 32.9(0.1) & 31.9(0.1) & 43.6(0.1) & 34.9(0.1) & 35.3(0.1) & 42.7(0.2) & 27.3(0.1) & 16.1(0.1) & 43.6(0.1) & 30.9(0.1) & 20.2(0.1) \\
\textbf{Min-Max Norm} & \textbf{BordaFuse} & 42.4(0.1) & 33.7(0.1) & 33.5(0.1) & 44.0(0.1) & 35.1(0.1) & 35.6(0.1) & 42.4(0.1) & 28.2(0.1) & 17.4(0.1) & 44.0(0.1) & 31.0(0.1) & 20.2(0.1) \\
\textbf{Min-Max Norm} & \textbf{Condorcet} & 36.7(0.2) & 27.4(0.1) & 26.8(0.1) & 38.7(0.2) & 29.6(0.2) & 29.8(0.1) & 36.7(0.2) & 22.4(0.1) & 13.6(0.1) & 38.7(0.2) & 25.7(0.1) & 16.7(0.1) \\
\hline

\textbf{ZMUV Norm} & \textbf{CombMNZ} & \textbf{47.3(0.1)} & \textbf{36.0(0.1)} & \textbf{35.1(0.1)} & \textbf{46.4(0.2)} & \textbf{37.0(0.1)} & \textbf{37.4(0.1)} & \textbf{47.3(0.1)} & \textbf{29.6(0.1)} & \textbf{17.7(0.0)} & \textbf{46.4(0.2)} & \textbf{32.7(0.2)} & \textbf{21.2(0.1)} \\
\textbf{ZMUV Norm} & \textbf{CombSUM} & \textbf{48.0(0.2)} & 35.5(0.1) & 34.6(0.1) & \textbf{46.1(0.2)} & 36.5(0.1) & 36.8(0.1) & \textbf{48.0(0.2)} & 28.9(0.1) & 17.3(0.1) & \textbf{46.1(0.2)} & 32.1(0.1) & 20.8(0.0) \\
\textbf{ZMUV Norm} & \textbf{CombMIN} & 31.7(0.1) & 25.3(0.1) & 25.9(0.1) & 32.7(0.2) & 27.1(0.1) & 27.9(0.1) & 31.7(0.1) & 21.3(0.1) & 14.0(0.1) & 32.7(0.2) & 24.1(0.1) & 16.3(0.1) \\
\textbf{ZMUV Norm} & \textbf{CombMAX} & 46.4(0.2) & 34.1(0.1) & 33.7(0.1) & 42.6(0.2) & 34.4(0.1) & 35.4(0.1) & 46.4(0.2) & 27.8(0.1) & 17.1(0.0) & 42.6(0.2) & 30.6(0.1) & 20.4(0.1) \\
\textbf{ZMUV Norm} & \textbf{CombMED} & 43.4(0.3) & 31.4(0.1) & 31.5(0.1) & 41.1(0.2) & 32.6(0.1) & 33.6(0.1) & 43.4(0.3) & 25.5(0.1) & 16.2(0.1) & 41.1(0.2) & 28.8(0.1) & 19.4(0.1) \\
\textbf{ZMUV Norm} & \textbf{CombANZ} & 43.4(0.3) & 31.4(0.1) & 31.5(0.1) & 41.1(0.2) & 32.6(0.1) & 33.6(0.1) & 43.4(0.3) & 25.5(0.1) & 16.2(0.1) & 41.1(0.2) & 28.8(0.1) & 19.4(0.1) \\
\textbf{ZMUV Norm} & \textbf{ISR} & 43.7(0.2) & 32.7(0.1) & 32.9(0.1) & 43.4(0.1) & 34.2(0.1) & 35.5(0.1) & 43.7(0.2) & 27.2(0.1) & 17.2(0.1) & 43.4(0.1) & 30.4(0.1) & 20.7(0.0) \\
\textbf{ZMUV Norm} & \textbf{Log\_ISR} & 42.7(0.2) & 32.9(0.1) & 31.9(0.1) & 43.6(0.1) & 34.9(0.1) & 35.3(0.1) & 42.7(0.2) & 27.3(0.1) & 16.1(0.1) & 43.6(0.1) & 30.9(0.1) & 20.2(0.1) \\
\textbf{ZMUV Norm} & \textbf{BordaFuse} & 42.4(0.1) & 33.7(0.1) & 33.5(0.1) & 44.0(0.1) & 35.1(0.1) & 35.6(0.1) & 42.4(0.1) & 28.2(0.1) & 17.4(0.1) & 44.0(0.1) & 31.0(0.1) & 20.2(0.1) \\
\textbf{ZMUV Norm} & \textbf{Condorcet} & 36.7(0.2) & 27.4(0.1) & 26.8(0.1) & 38.7(0.2) & 29.6(0.2) & 29.8(0.1) & 36.7(0.2) & 22.4(0.1) & 13.6(0.1) & 38.7(0.2) & 25.7(0.1) & 16.7(0.1) \\
\hline

\textbf{Max Norm} & \textbf{CombMNZ} & 43.4(0.2) & 33.9(0.1) & 33.3(0.0) & 45.3(0.2) & 36.4(0.1) & 36.6(0.1) & 43.4(0.2) & 28.2(0.1) & 17.0(0.1) & 45.3(0.2) & 32.1(0.1) & 20.8(0.1) \\
\textbf{Max Norm} & \textbf{CombSUM} & 43.4(0.2) & 33.9(0.1) & 33.3(0.1) & 45.3(0.2) & 36.4(0.1) & 36.7(0.1) & 43.4(0.2) & 28.2(0.1) & 17.1(0.1) & 45.3(0.2) & 32.2(0.1) & 20.8(0.1) \\
\textbf{Max Norm} & \textbf{CombMIN} & 25.0(0.2) & 21.2(0.2) & 22.2(0.2) & 33.0(0.1) & 26.1(0.1) & 26.9(0.1) & 25.0(0.2) & 18.3(0.1) & 12.4(0.1) & 33.0(0.1) & 23.0(0.1) & 15.5(0.1) \\
\textbf{Max Norm} & \textbf{CombMAX} & 28.3(0.1) & 27.1(0.1) & 28.1(0.1) & 36.3(0.2) & 32.8(0.1) & 33.7(0.1) & 28.3(0.1) & 24.1(0.1) & 15.9(0.0) & 36.3(0.2) & 30.1(0.2) & 19.9(0.1) \\
\textbf{Max Norm} & \textbf{CombMED} & 25.5(0.2) & 23.5(0.1) & 24.9(0.1) & 34.1(0.2) & 29.6(0.1) & 30.9(0.1) & 25.5(0.2) & 21.0(0.1) & 14.3(0.1) & 34.1(0.2) & 27.0(0.1) & 18.3(0.1) \\
\textbf{Max Norm} & \textbf{CombANZ} & 25.5(0.2) & 23.5(0.1) & 24.9(0.1) & 34.1(0.2) & 29.6(0.1) & 30.9(0.1) & 25.5(0.2) & 21.0(0.1) & 14.3(0.1) & 34.1(0.2) & 27.0(0.1) & 18.3(0.1) \\
\textbf{Max Norm} & \textbf{ISR} & 43.7(0.2) & 32.7(0.1) & 32.9(0.1) & 43.4(0.1) & 34.2(0.1) & 35.5(0.1) & 43.7(0.2) & 27.2(0.1) & 17.2(0.1) & 43.4(0.1) & 30.4(0.1) & 20.7(0.0) \\
\textbf{Max Norm} & \textbf{Log\_ISR} & 42.7(0.2) & 32.9(0.1) & 31.9(0.1) & 43.6(0.1) & 34.9(0.1) & 35.3(0.1) & 42.7(0.2) & 27.3(0.1) & 16.1(0.1) & 43.6(0.1) & 30.9(0.1) & 20.2(0.1) \\
\textbf{Max Norm} & \textbf{BordaFuse} & 42.4(0.1) & 33.7(0.1) & 33.5(0.1) & 44.0(0.1) & 35.1(0.1) & 35.6(0.1) & 42.4(0.1) & 28.2(0.1) & 17.4(0.1) & 44.0(0.1) & 31.0(0.1) & 20.2(0.1) \\
\textbf{Max Norm} & \textbf{Condorcet} & 36.7(0.2) & 27.4(0.1) & 26.8(0.1) & 38.7(0.2) & 29.6(0.2) & 29.8(0.1) & 36.7(0.2) & 22.4(0.1) & 13.6(0.1) & 38.7(0.2) & 25.7(0.1) & 16.7(0.1) \\
\hline

\textbf{Sum Norm} & \textbf{CombMNZ} & 45.7(0.1) & 35.2(0.1) & 34.5(0.0) & \textbf{46.2(0.1)} & \textbf{37.2(0.1)} & \textbf{37.5(0.1)} & 45.7(0.1) & 29.1(0.1) & 17.5(0.1) & \textbf{46.2(0.1)} & \textbf{32.9(0.1)} & \textbf{21.3(0.1)} \\
\textbf{Sum Norm} & \textbf{CombSUM} & 46.5(0.1) & 34.9(0.1) & 34.2(0.1) & \textbf{46.2(0.1)} & 36.9(0.1) & 37.2(0.1) & 46.5(0.1) & 28.6(0.1) & 17.3(0.1) & \textbf{46.2(0.1)} & 32.6(0.1) & 21.1(0.0) \\
\textbf{Sum Norm} & \textbf{CombMIN} & 27.2(0.3) & 23.2(0.2) & 24.3(0.1) & 31.2(0.2) & 25.8(0.2) & 26.9(0.1) & 27.2(0.3) & 20.0(0.1) & 13.7(0.1) & 31.2(0.2) & 23.0(0.1) & 15.8(0.1) \\
\textbf{Sum Norm} & \textbf{CombMAX} & 42.3(0.1) & 31.9(0.1) & 32.0(0.1) & 42.3(0.2) & 34.9(0.1) & 35.5(0.1) & 42.3(0.1) & 26.4(0.1) & 16.7(0.1) & 42.3(0.2) & 31.3(0.1) & 20.4(0.1) \\
\textbf{Sum Norm} & \textbf{CombMED} & 36.5(0.3) & 28.7(0.2) & 29.4(0.2) & 38.5(0.2) & 32.3(0.1) & 33.4(0.1) & 36.5(0.3) & 24.2(0.1) & 15.9(0.1) & 38.5(0.2) & 29.1(0.1) & 19.5(0.1) \\
\textbf{Sum Norm} & \textbf{CombANZ} & 36.5(0.3) & 28.7(0.2) & 29.4(0.2) & 38.5(0.2) & 32.3(0.1) & 33.4(0.1) & 36.5(0.3) & 24.2(0.1) & 15.9(0.1) & 38.5(0.2) & 29.1(0.1) & 19.5(0.1) \\
\textbf{Sum Norm} & \textbf{ISR} & 43.7(0.2) & 32.7(0.1) & 32.9(0.1) & 43.4(0.1) & 34.2(0.1) & 35.5(0.1) & 43.7(0.2) & 27.2(0.1) & 17.2(0.1) & 43.4(0.1) & 30.4(0.1) & 20.7(0.0) \\
\textbf{Sum Norm} & \textbf{Log\_ISR} & 42.7(0.2) & 32.9(0.1) & 31.9(0.1) & 43.6(0.1) & 34.9(0.1) & 35.3(0.1) & 42.7(0.2) & 27.3(0.1) & 16.1(0.1) & 43.6(0.1) & 30.9(0.1) & 20.2(0.1) \\
\textbf{Sum Norm} & \textbf{BordaFuse} & 42.4(0.1) & 33.7(0.1) & 33.5(0.1) & 44.0(0.1) & 35.1(0.1) & 35.6(0.1) & 42.4(0.1) & 28.2(0.1) & 17.4(0.1) & 44.0(0.1) & 31.0(0.1) & 20.2(0.1) \\
\textbf{Sum Norm} & \textbf{Condorcet} & 36.7(0.2) & 27.4(0.1) & 26.8(0.1) & 38.7(0.2) & 29.6(0.2) & 29.8(0.1) & 36.7(0.2) & 22.4(0.1) & 13.6(0.1) & 38.7(0.2) & 25.7(0.1) & 16.7(0.1) \\
\hline

\textbf{Rank Norm} & \textbf{CombMNZ} & 42.0(0.1) & 33.4(0.1) & 33.2(0.0) & 43.8(0.1) & 35.3(0.1) & 35.9(0.1) & 42.0(0.1) & 27.9(0.1) & 17.3(0.0) & 43.8(0.1) & 31.3(0.1) & 20.5(0.1) \\
\textbf{Rank Norm} & \textbf{CombSUM} & 42.1(0.2) & 33.6(0.1) & 33.5(0.1) & 43.9(0.1) & 35.4(0.1) & 36.1(0.1) & 42.1(0.2) & 28.1(0.1) & 17.5(0.1) & 43.9(0.1) & 31.4(0.1) & 20.7(0.1) \\
\textbf{Rank Norm} & \textbf{CombMIN} & 30.9(0.2) & 23.9(0.1) & 24.8(0.1) & 33.0(0.1) & 25.9(0.1) & 26.9(0.1) & 30.9(0.2) & 20.1(0.1) & 13.5(0.1) & 33.0(0.1) & 22.7(0.1) & 15.6(0.1) \\
\textbf{Rank Norm} & \textbf{CombMAX} & 31.5(0.1) & 30.5(0.1) & 31.0(0.1) & 36.3(0.1) & 32.0(0.1) & 33.6(0.1) & 31.5(0.1) & 26.7(0.1) & 17.0(0.1) & 36.3(0.1) & 28.8(0.1) & 20.0(0.1) \\
\textbf{Rank Norm} & \textbf{CombMED} & 32.2(0.2) & 26.0(0.1) & 26.7(0.1) & 34.7(0.1) & 28.1(0.1) & 29.2(0.1) & 32.2(0.2) & 22.2(0.1) & 14.6(0.1) & 34.7(0.1) & 25.0(0.1) & 17.1(0.1) \\
\textbf{Rank Norm} & \textbf{CombANZ} & 32.2(0.2) & 26.0(0.1) & 26.7(0.1) & 34.7(0.1) & 28.1(0.1) & 29.2(0.1) & 32.2(0.2) & 22.2(0.1) & 14.6(0.1) & 34.7(0.1) & 25.0(0.1) & 17.1(0.1) \\
\textbf{Rank Norm} & \textbf{ISR} & 43.7(0.2) & 32.7(0.1) & 32.9(0.1) & 43.4(0.1) & 34.2(0.1) & 35.5(0.1) & 43.7(0.2) & 27.2(0.1) & 17.2(0.1) & 43.4(0.1) & 30.4(0.1) & 20.7(0.0) \\
\textbf{Rank Norm} & \textbf{Log\_ISR} & 42.7(0.2) & 32.9(0.1) & 31.9(0.1) & 43.6(0.1) & 34.9(0.1) & 35.3(0.1) & 42.7(0.2) & 27.3(0.1) & 16.1(0.1) & 43.6(0.1) & 30.9(0.1) & 20.2(0.1) \\
\textbf{Rank Norm} & \textbf{BordaFuse} & 42.4(0.1) & 33.7(0.1) & 33.5(0.1) & 44.0(0.1) & 35.1(0.1) & 35.6(0.1) & 42.4(0.1) & 28.2(0.1) & 17.4(0.1) & 44.0(0.1) & 31.0(0.1) & 20.2(0.1) \\
\textbf{Rank Norm} & \textbf{Condorcet} & 36.7(0.2) & 27.4(0.1) & 26.8(0.1) & 38.7(0.2) & 29.6(0.2) & 29.8(0.1) & 36.7(0.2) & 22.4(0.1) & 13.6(0.1) & 38.7(0.2) & 25.7(0.1) & 16.7(0.1) \\
\hline

\textbf{Borda Norm} & \textbf{CombMNZ} & 42.5(0.2) & 33.8(0.1) & 33.5(0.1) & 44.0(0.1) & 35.1(0.1) & 35.6(0.2) & 42.5(0.2) & 28.2(0.1) & 17.4(0.1) & 44.0(0.1) & 30.9(0.2) & 20.2(0.1) \\
\textbf{Borda Norm} & \textbf{CombSUM} & 42.5(0.2) & 33.8(0.1) & 33.5(0.1) & 44.0(0.1) & 35.1(0.1) & 35.6(0.2) & 42.5(0.2) & 28.2(0.1) & 17.4(0.1) & 44.0(0.1) & 30.9(0.2) & 20.2(0.1) \\
\textbf{Borda Norm} & \textbf{CombMIN} & 40.9(0.1) & 31.9(0.1) & 31.1(0.1) & 43.0(0.1) & 33.9(0.1) & 34.0(0.2) & 40.9(0.1) & 26.5(0.1) & 15.8(0.1) & 43.0(0.1) & 29.7(0.1) & 19.2(0.1) \\
\textbf{Borda Norm} & \textbf{CombMAX} & 36.9(0.1) & 30.8(0.1) & 31.3(0.1) & 38.4(0.2) & 32.9(0.1) & 34.3(0.1) & 36.9(0.1) & 26.1(0.1) & 16.8(0.1) & 38.4(0.2) & 29.6(0.1) & 20.3(0.1) \\
\textbf{Borda Norm} & \textbf{CombMED} & 42.5(0.2) & 33.8(0.1) & 33.5(0.1) & 44.0(0.1) & 35.1(0.1) & 35.6(0.2) & 42.5(0.2) & 28.2(0.1) & 17.4(0.1) & 44.0(0.1) & 30.9(0.2) & 20.2(0.1) \\
\textbf{Borda Norm} & \textbf{CombANZ} & 42.5(0.2) & 33.8(0.1) & 33.5(0.1) & 44.0(0.1) & 35.1(0.1) & 35.6(0.2) & 42.5(0.2) & 28.2(0.1) & 17.4(0.1) & 44.0(0.1) & 30.9(0.2) & 20.2(0.1) \\
\textbf{Borda Norm} & \textbf{ISR} & 1.7(0.0) & 1.7(0.0) & 2.3(0.0) & 2.4(0.1) & 2.2(0.1) & 3.0(0.1) & 1.7(0.0) & 1.6(0.0) & 1.5(0.1) & 2.4(0.1) & 2.2(0.1) & 2.1(0.0) \\
\textbf{Borda Norm} & \textbf{Log\_ISR} & 1.7(0.0) & 1.7(0.0) & 2.3(0.0) & 2.4(0.1) & 2.2(0.1) & 3.0(0.1) & 1.7(0.0) & 1.6(0.0) & 1.5(0.1) & 2.4(0.1) & 2.2(0.1) & 2.1(0.0) \\
\textbf{Borda Norm} & \textbf{BordaFuse} & 1.7(0.0) & 1.7(0.0) & 2.3(0.0) & 2.4(0.1) & 2.2(0.1) & 3.0(0.1) & 1.7(0.0) & 1.6(0.0) & 1.5(0.1) & 2.4(0.1) & 2.2(0.1) & 2.1(0.0) \\
\textbf{Borda Norm} & \textbf{Condorcet} & 1.7(0.0) & 1.7(0.0) & 2.3(0.0) & 2.4(0.1) & 2.2(0.1) & 3.0(0.1) & 1.7(0.0) & 1.6(0.0) & 1.5(0.1) & 2.4(0.1) & 2.2(0.1) & 2.1(0.0) \\

\hline

\end{tabular}
}
\end{table*}

\begin{table*}
\centering
\caption{nDCG@\textit{k} and Precision@\textit{k} of Normalization Strategies and Fusion Algorithms on \textbf{Amazon-13k}.}
\label{tb_norm_fuse_amazoncat_13k}

\resizebox{\textwidth}{!}{

\begin{tabular}{l l | c c c | c c c | c c c | c c c}

\hline

 &  & \multicolumn{6}{c|}{\textbf{nDCG x 100}} & \multicolumn{6}{c}{\textbf{Precision x 100}} \\
 &  & \multicolumn{3}{c|}{\textbf{Tail label}} & \multicolumn{3}{c|}{\textbf{Head label}} & \multicolumn{3}{c|}{\textbf{Tail label}} & \multicolumn{3}{c}{\textbf{Head label}} \\

\textbf{Normalization } & \textbf{Algorithms} & \textbf{@1} & \textbf{@5} & \textbf{@10} & \textbf{@1} & \textbf{@5} & \textbf{@10} & \textbf{@1} & \textbf{@5} & \textbf{@10} & \textbf{@1} & \textbf{@5} & \textbf{@10} \\
\hline

\textbf{Min-Max Norm} & \textbf{CombMNZ} & 70.5(0.2) & \textbf{34.0(0.1)} & \textbf{31.2(0.1)} & \textbf{95.6(0.1)} & 86.5(0.1) & 86.5(0.1) & 70.5(0.2) & \textbf{22.7(0.1)} & \textbf{12.0(0.1)} & \textbf{95.6(0.1)} & 63.7(0.1) & 38.5(0.0) \\
\textbf{Min-Max Norm} & \textbf{CombSUM} & 69.9(0.2) & 33.8(0.1) & \textbf{31.0(0.1)} & \textbf{95.6(0.1)} & 86.0(0.1) & 85.9(0.1) & 69.9(0.2) & \textbf{22.7(0.1)} & \textbf{12.0(0.1)} & \textbf{95.6(0.1)} & 63.2(0.1) & 38.0(0.1) \\
\textbf{Min-Max Norm} & \textbf{CombMIN} & 49.2(0.6) & 25.2(0.1) & 23.9(0.1) & 82.8(0.1) & 69.4(0.2) & 71.2(0.1) & 49.2(0.6) & 17.6(0.1) & 9.9(0.0) & 82.8(0.1) & 49.2(0.2) & 30.7(0.1) \\
\textbf{Min-Max Norm} & \textbf{CombMAX} & 61.4(0.1) & 31.8(0.1) & 29.4(0.1) & 74.3(0.0) & 74.3(0.1) & 76.9(0.0) & 61.4(0.1) & 22.0(0.1) & 11.8(0.0) & 74.3(0.0) & 56.5(0.0) & 35.9(0.1) \\
\textbf{Min-Max Norm} & \textbf{CombMED} & 52.3(0.7) & 28.5(0.1) & 26.8(0.1) & 85.0(0.1) & 76.1(0.1) & 78.2(0.1) & 52.3(0.7) & 20.3(0.1) & 11.3(0.0) & 85.0(0.1) & 55.6(0.1) & 35.0(0.1) \\
\textbf{Min-Max Norm} & \textbf{CombANZ} & 52.3(0.7) & 28.5(0.1) & 26.8(0.1) & 85.0(0.1) & 76.1(0.1) & 78.2(0.1) & 52.3(0.7) & 20.3(0.1) & 11.3(0.0) & 85.0(0.1) & 55.6(0.1) & 35.0(0.1) \\
\textbf{Min-Max Norm} & \textbf{ISR} & 67.6(0.1) & 33.0(0.1) & 30.4(0.1) & 92.8(0.1) & 83.4(0.1) & 84.6(0.1) & 67.6(0.1) & 22.4(0.1) & \textbf{12.0(0.1)} & 92.8(0.1) & 61.8(0.0) & 38.3(0.1) \\
\textbf{Min-Max Norm} & \textbf{Log\_ISR} & 68.2(0.1) & 33.1(0.1) & 30.3(0.1) & 92.8(0.1) & 84.4(0.1) & 85.2(0.1) & 68.2(0.1) & 22.2(0.1) & 11.7(0.1) & 92.8(0.1) & 62.6(0.1) & 38.6(0.1) \\
\textbf{Min-Max Norm} & \textbf{BordaFuse} & 67.9(0.2) & 33.1(0.1) & 30.4(0.0) & 91.9(0.1) & 85.3(0.1) & 85.7(0.0) & 67.9(0.2) & 22.3(0.1) & \textbf{11.9(0.1)} & 91.9(0.1) & 63.6(0.1) & \textbf{38.8(0.0)} \\
\textbf{Min-Max Norm} & \textbf{Condorcet} & 57.6(0.4) & 29.1(0.1) & 27.2(0.1) & 86.0(1.7) & 78.5(0.6) & 79.2(0.5) & 57.6(0.4) & 20.0(0.1) & 11.0(0.1) & 86.0(1.7) & 57.9(0.3) & 35.3(0.1) \\
\hline

\textbf{ZMUV Norm} & \textbf{CombMNZ} & \textbf{71.2(0.2)} & \textbf{34.1(0.1)} & \textbf{31.2(0.1)} & \textbf{95.7(0.1)} & \textbf{87.4(0.1)} & \textbf{87.2(0.1)} & \textbf{71.2(0.2)} & \textbf{22.8(0.1)} & \textbf{12.0(0.1)} & \textbf{95.7(0.1)} & \textbf{64.5(0.1)} & \textbf{38.9(0.1)} \\
\textbf{ZMUV Norm} & \textbf{CombSUM} & \textbf{69.0(0.4)} & 33.1(0.1) & 30.4(0.1) & \textbf{95.6(0.1)} & 87.0(0.1) & 86.7(0.1) & \textbf{69.0(0.4)} & 22.1(0.1) & \textbf{11.8(0.1)} & \textbf{95.6(0.1)} & 64.1(0.1) & 38.5(0.1) \\
\textbf{ZMUV Norm} & \textbf{CombMIN} & 22.2(0.6) & 16.5(0.1) & 17.0(0.1) & 63.4(0.2) & 63.8(0.2) & 66.9(0.1) & 22.2(0.6) & 13.6(0.0) & 8.9(0.0) & 63.4(0.2) & 48.4(0.1) & 32.0(0.1) \\
\textbf{ZMUV Norm} & \textbf{CombMAX} & 56.2(0.7) & 29.8(0.1) & 27.9(0.1) & 94.2(0.2) & 84.7(0.2) & 85.1(0.1) & 56.2(0.7) & 21.1(0.1) & 11.7(0.0) & 94.2(0.2) & 62.0(0.1) & 37.9(0.1) \\
\textbf{ZMUV Norm} & \textbf{CombMED} & 44.4(0.7) & 25.9(0.1) & 24.8(0.1) & 92.7(0.2) & 82.6(0.1) & 83.3(0.1) & 44.4(0.7) & 19.2(0.1) & 11.1(0.0) & 92.7(0.2) & 60.2(0.1) & 37.0(0.1) \\
\textbf{ZMUV Norm} & \textbf{CombANZ} & 44.4(0.7) & 25.9(0.1) & 24.8(0.1) & 92.7(0.2) & 82.6(0.1) & 83.3(0.1) & 44.4(0.7) & 19.2(0.1) & 11.1(0.0) & 92.7(0.2) & 60.2(0.1) & 37.0(0.1) \\
\textbf{ZMUV Norm} & \textbf{ISR} & 67.6(0.1) & 33.0(0.1) & 30.4(0.1) & 92.8(0.1) & 83.4(0.1) & 84.6(0.1) & 67.6(0.1) & 22.4(0.1) & \textbf{12.0(0.1)} & 92.8(0.1) & 61.8(0.0) & 38.3(0.1) \\
\textbf{ZMUV Norm} & \textbf{Log\_ISR} & 68.2(0.1) & 33.1(0.1) & 30.3(0.1) & 92.8(0.1) & 84.4(0.1) & 85.2(0.1) & 68.2(0.1) & 22.2(0.1) & 11.7(0.1) & 92.8(0.1) & 62.6(0.1) & 38.6(0.1) \\
\textbf{ZMUV Norm} & \textbf{BordaFuse} & 67.9(0.2) & 33.1(0.1) & 30.4(0.0) & 91.9(0.1) & 85.3(0.1) & 85.7(0.0) & 67.9(0.2) & 22.3(0.1) & \textbf{11.9(0.1)} & 91.9(0.1) & 63.6(0.1) & \textbf{38.8(0.0)} \\
\textbf{ZMUV Norm} & \textbf{Condorcet} & 57.6(0.4) & 29.1(0.1) & 27.2(0.1) & 86.0(1.7) & 78.5(0.6) & 79.2(0.5) & 57.6(0.4) & 20.0(0.1) & 11.0(0.1) & 86.0(1.7) & 57.9(0.3) & 35.3(0.1) \\
\hline

\textbf{Max Norm} & \textbf{CombMNZ} & 69.4(0.1) & 33.5(0.1) & 30.8(0.1) & 95.5(0.1) & 86.9(0.1) & 86.8(0.1) & 69.4(0.1) & 22.4(0.1) & \textbf{11.9(0.1)} & 95.5(0.1) & 64.0(0.0) & 38.6(0.1) \\
\textbf{Max Norm} & \textbf{CombSUM} & 69.5(0.2) & 33.5(0.1) & 30.8(0.1) & 95.5(0.1) & 86.9(0.1) & 86.8(0.1) & 69.5(0.2) & 22.4(0.1) & \textbf{11.9(0.1)} & 95.5(0.1) & 64.0(0.0) & 38.6(0.1) \\
\textbf{Max Norm} & \textbf{CombMIN} & 50.2(0.6) & 25.3(0.1) & 23.7(0.1) & 85.7(0.1) & 70.8(0.2) & 71.8(0.2) & 50.2(0.6) & 17.3(0.1) & 9.6(0.0) & 85.7(0.1) & 50.1(0.3) & 30.5(0.2) \\
\textbf{Max Norm} & \textbf{CombMAX} & 61.9(0.1) & 31.8(0.1) & 29.4(0.1) & 74.3(0.1) & 74.4(0.1) & 76.8(0.1) & 61.9(0.1) & 21.9(0.1) & \textbf{11.8(0.1)} & 74.3(0.1) & 56.5(0.1) & 35.8(0.1) \\
\textbf{Max Norm} & \textbf{CombMED} & 52.9(0.6) & 27.9(0.1) & 26.0(0.1) & 86.1(0.1) & 75.6(0.2) & 76.7(0.2) & 52.9(0.6) & 19.5(0.1) & 10.5(0.1) & 86.1(0.1) & 54.8(0.2) & 33.6(0.1) \\
\textbf{Max Norm} & \textbf{CombANZ} & 52.9(0.6) & 27.9(0.1) & 26.0(0.1) & 86.1(0.1) & 75.6(0.2) & 76.7(0.2) & 52.9(0.6) & 19.5(0.1) & 10.5(0.1) & 86.1(0.1) & 54.8(0.2) & 33.6(0.1) \\
\textbf{Max Norm} & \textbf{ISR} & 67.6(0.1) & 33.0(0.1) & 30.4(0.1) & 92.8(0.1) & 83.4(0.1) & 84.6(0.1) & 67.6(0.1) & 22.4(0.1) & \textbf{12.0(0.1)} & 92.8(0.1) & 61.8(0.0) & 38.3(0.1) \\
\textbf{Max Norm} & \textbf{Log\_ISR} & 68.2(0.1) & 33.1(0.1) & 30.3(0.1) & 92.8(0.1) & 84.4(0.1) & 85.2(0.1) & 68.2(0.1) & 22.2(0.1) & 11.7(0.1) & 92.8(0.1) & 62.6(0.1) & 38.6(0.1) \\
\textbf{Max Norm} & \textbf{BordaFuse} & 67.9(0.2) & 33.1(0.1) & 30.4(0.0) & 91.9(0.1) & 85.3(0.1) & 85.7(0.0) & 67.9(0.2) & 22.3(0.1) & \textbf{11.9(0.1)} & 91.9(0.1) & 63.6(0.1) & \textbf{38.8(0.0)} \\
\textbf{Max Norm} & \textbf{Condorcet} & 57.6(0.4) & 29.1(0.1) & 27.2(0.1) & 86.0(1.7) & 78.5(0.6) & 79.2(0.5) & 57.6(0.4) & 20.0(0.1) & 11.0(0.1) & 86.0(1.7) & 57.9(0.3) & 35.3(0.1) \\
\hline

\textbf{Sum Norm} & \textbf{CombMNZ} & 68.0(0.1) & 33.1(0.1) & 30.5(0.1) & \textbf{95.7(0.1)} & \textbf{87.2(0.1)} & \textbf{87.2(0.1)} & 68.0(0.1) & 22.2(0.1) & \textbf{11.9(0.1)} & \textbf{95.7(0.1)} & \textbf{64.2(0.1)} & \textbf{38.9(0.1)} \\
\textbf{Sum Norm} & \textbf{CombSUM} & 66.7(0.2) & 32.6(0.1) & 30.2(0.1) & \textbf{95.6(0.1)} & 87.0(0.1) & 87.0(0.1) & 66.7(0.2) & 22.0(0.1) & \textbf{11.8(0.1)} & \textbf{95.6(0.1)} & 64.1(0.0) & \textbf{38.8(0.1)} \\
\textbf{Sum Norm} & \textbf{CombMIN} & 31.2(0.5) & 21.3(0.1) & 21.0(0.1) & 80.1(0.2) & 73.0(0.1) & 74.6(0.1) & 31.2(0.5) & 17.0(0.1) & 10.2(0.0) & 80.1(0.2) & 53.7(0.1) & 33.7(0.1) \\
\textbf{Sum Norm} & \textbf{CombMAX} & 62.6(0.1) & 31.4(0.1) & 29.3(0.1) & 90.4(0.2) & 82.2(0.1) & 83.2(0.1) & 62.6(0.1) & 21.4(0.1) & 11.7(0.1) & 90.4(0.2) & 60.6(0.1) & 37.4(0.0) \\
\textbf{Sum Norm} & \textbf{CombMED} & 58.9(0.2) & 30.3(0.1) & 28.4(0.1) & 92.3(0.1) & 82.5(0.1) & 83.3(0.1) & 58.9(0.2) & 21.0(0.1) & 11.6(0.1) & 92.3(0.1) & 60.1(0.1) & 37.0(0.0) \\
\textbf{Sum Norm} & \textbf{CombANZ} & 58.9(0.2) & 30.3(0.1) & 28.4(0.1) & 92.3(0.1) & 82.5(0.1) & 83.3(0.1) & 58.9(0.2) & 21.0(0.1) & 11.6(0.1) & 92.3(0.1) & 60.1(0.1) & 37.0(0.0) \\
\textbf{Sum Norm} & \textbf{ISR} & 67.6(0.1) & 33.0(0.1) & 30.4(0.1) & 92.8(0.1) & 83.4(0.1) & 84.6(0.1) & 67.6(0.1) & 22.4(0.1) & \textbf{12.0(0.1)} & 92.8(0.1) & 61.8(0.0) & 38.3(0.1) \\
\textbf{Sum Norm} & \textbf{Log\_ISR} & 68.2(0.1) & 33.1(0.1) & 30.3(0.1) & 92.8(0.1) & 84.4(0.1) & 85.2(0.1) & 68.2(0.1) & 22.2(0.1) & 11.7(0.1) & 92.8(0.1) & 62.6(0.1) & 38.6(0.1) \\
\textbf{Sum Norm} & \textbf{BordaFuse} & 67.9(0.2) & 33.1(0.1) & 30.4(0.0) & 91.9(0.1) & 85.3(0.1) & 85.7(0.0) & 67.9(0.2) & 22.3(0.1) & \textbf{11.9(0.1)} & 91.9(0.1) & 63.6(0.1) & \textbf{38.8(0.0)} \\
\textbf{Sum Norm} & \textbf{Condorcet} & 57.6(0.4) & 29.1(0.1) & 27.2(0.1) & 86.0(1.7) & 78.5(0.6) & 79.2(0.5) & 57.6(0.4) & 20.0(0.1) & 11.0(0.1) & 86.0(1.7) & 57.9(0.3) & 35.3(0.1) \\
\hline

\textbf{Rank Norm} & \textbf{CombMNZ} & 67.6(0.1) & 33.1(0.1) & 30.5(0.1) & 88.8(0.1) & 83.6(0.1) & 84.3(0.1) & 67.6(0.1) & 22.4(0.1) & \textbf{12.0(0.1)} & 88.8(0.1) & 62.7(0.1) & 38.6(0.0) \\
\textbf{Rank Norm} & \textbf{CombSUM} & 67.6(0.1) & 33.2(0.1) & 30.6(0.1) & 88.8(0.1) & 83.6(0.1) & 84.4(0.1) & 67.6(0.1) & 22.4(0.1) & \textbf{12.0(0.1)} & 88.8(0.1) & 62.7(0.1) & 38.6(0.0) \\
\textbf{Rank Norm} & \textbf{CombMIN} & 45.8(0.6) & 21.9(0.1) & 21.5(0.1) & 78.9(0.1) & 61.6(0.1) & 64.5(0.1) & 45.8(0.6) & 14.9(0.1) & 9.1(0.1) & 78.9(0.1) & 43.1(0.1) & 28.8(0.1) \\
\textbf{Rank Norm} & \textbf{CombMAX} & 61.5(0.1) & 31.1(0.1) & 29.0(0.1) & 75.3(0.0) & 80.1(0.1) & 81.3(0.1) & 61.5(0.1) & 21.3(0.1) & 11.7(0.0) & 75.3(0.0) & 61.3(0.1) & 37.9(0.0) \\
\textbf{Rank Norm} & \textbf{CombMED} & 48.3(0.7) & 24.4(0.1) & 23.7(0.1) & 81.1(0.1) & 68.6(0.1) & 71.6(0.1) & 48.3(0.7) & 17.0(0.0) & 10.1(0.0) & 81.1(0.1) & 49.5(0.1) & 32.8(0.1) \\
\textbf{Rank Norm} & \textbf{CombANZ} & 48.3(0.7) & 24.4(0.1) & 23.7(0.1) & 81.1(0.1) & 68.6(0.1) & 71.6(0.1) & 48.3(0.7) & 17.0(0.0) & 10.1(0.0) & 81.1(0.1) & 49.5(0.1) & 32.8(0.1) \\
\textbf{Rank Norm} & \textbf{ISR} & 67.6(0.1) & 33.0(0.1) & 30.4(0.1) & 92.8(0.1) & 83.4(0.1) & 84.6(0.1) & 67.6(0.1) & 22.4(0.1) & \textbf{12.0(0.1)} & 92.8(0.1) & 61.8(0.0) & 38.3(0.1) \\
\textbf{Rank Norm} & \textbf{Log\_ISR} & 68.2(0.1) & 33.1(0.1) & 30.3(0.1) & 92.8(0.1) & 84.4(0.1) & 85.2(0.1) & 68.2(0.1) & 22.2(0.1) & 11.7(0.1) & 92.8(0.1) & 62.6(0.1) & 38.6(0.1) \\
\textbf{Rank Norm} & \textbf{BordaFuse} & 67.9(0.2) & 33.1(0.1) & 30.4(0.0) & 91.9(0.1) & 85.3(0.1) & 85.7(0.0) & 67.9(0.2) & 22.3(0.1) & \textbf{11.9(0.1)} & 91.9(0.1) & 63.6(0.1) & \textbf{38.8(0.0)} \\
\textbf{Rank Norm} & \textbf{Condorcet} & 57.6(0.4) & 29.1(0.1) & 27.2(0.1) & 86.0(1.7) & 78.5(0.6) & 79.2(0.5) & 57.6(0.4) & 20.0(0.1) & 11.0(0.1) & 86.0(1.7) & 57.9(0.3) & 35.3(0.1) \\
\hline

\textbf{Borda Norm} & \textbf{CombMNZ} & 67.2(0.2) & 33.0(0.1) & 30.3(0.1) & 92.6(0.0) & 85.5(0.1) & 85.9(0.0) & 67.2(0.2) & 22.3(0.1) & \textbf{11.9(0.1)} & 92.6(0.0) & 63.6(0.0) & \textbf{38.8(0.0)} \\
\textbf{Borda Norm} & \textbf{CombSUM} & 67.2(0.2) & 33.0(0.1) & 30.3(0.1) & 92.6(0.0) & 85.5(0.1) & 85.9(0.0) & 67.2(0.2) & 22.3(0.1) & \textbf{11.9(0.1)} & 92.6(0.0) & 63.6(0.0) & \textbf{38.8(0.0)} \\
\textbf{Borda Norm} & \textbf{CombMIN} & 65.8(0.2) & 32.1(0.1) & 29.4(0.1) & 91.7(0.1) & 84.4(0.1) & 84.9(0.1) & 65.8(0.2) & 21.6(0.1) & 11.4(0.1) & 91.7(0.1) & 62.7(0.1) & 38.3(0.1) \\
\textbf{Borda Norm} & \textbf{CombMAX} & 57.9(0.4) & 31.3(0.1) & 28.9(0.1) & 84.7(1.4) & 81.1(0.2) & 82.5(0.2) & 57.9(0.4) & 22.1(0.1) & \textbf{11.9(0.0)} & 84.7(1.4) & 60.7(0.1) & 37.9(0.1) \\
\textbf{Borda Norm} & \textbf{CombMED} & 67.2(0.2) & 33.0(0.1) & 30.3(0.1) & 92.6(0.0) & 85.5(0.1) & 85.9(0.0) & 67.2(0.2) & 22.3(0.1) & \textbf{11.9(0.1)} & 92.6(0.0) & 63.6(0.0) & \textbf{38.8(0.0)} \\
\textbf{Borda Norm} & \textbf{CombANZ} & 67.2(0.2) & 33.0(0.1) & 30.3(0.1) & 92.6(0.0) & 85.5(0.1) & 85.9(0.0) & 67.2(0.2) & 22.3(0.1) & \textbf{11.9(0.1)} & 92.6(0.0) & 63.6(0.0) & \textbf{38.8(0.0)} \\
\textbf{Borda Norm} & \textbf{ISR} & 1.3(0.1) & 1.3(0.1) & 1.8(0.1) & 2.9(0.9) & 4.3(1.5) & 6.0(1.8) & 1.3(0.1) & 1.3(0.1) & 1.3(0.1) & 2.9(0.9) & 4.0(1.1) & 4.1(0.8) \\
\textbf{Borda Norm} & \textbf{Log\_ISR} & 1.3(0.1) & 1.3(0.1) & 1.8(0.1) & 2.9(0.9) & 4.3(1.5) & 6.0(1.8) & 1.3(0.1) & 1.3(0.1) & 1.3(0.1) & 2.9(0.9) & 4.0(1.1) & 4.1(0.8) \\
\textbf{Borda Norm} & \textbf{BordaFuse} & 1.3(0.1) & 1.3(0.1) & 1.8(0.1) & 2.9(0.9) & 4.3(1.5) & 6.0(1.8) & 1.3(0.1) & 1.3(0.1) & 1.3(0.1) & 2.9(0.9) & 4.0(1.1) & 4.1(0.8) \\
\textbf{Borda Norm} & \textbf{Condorcet} & 1.3(0.1) & 1.3(0.1) & 1.8(0.1) & 2.9(0.9) & 4.3(1.5) & 6.0(1.8) & 1.3(0.1) & 1.3(0.1) & 1.3(0.1) & 2.9(0.9) & 4.0(1.1) & 4.1(0.8) \\

\hline

\end{tabular}
}
\end{table*}

\section{Conclusion}
\label{sec_conclusion_and_future_work}

Employing the CombMNZ ranking-based fusion algorithm in conjunction with the ZMUV normalization strategy for fusing both dense and sparse rankings yielded the highest effectiveness across all datasets.

\section*{Acknowledgments}
This work was partially supported by CNPq, Capes, Fapemig, Fapesp, Google, AWS, and 
 Instituto Nacional de Ciência e Tecnologia em Inteligência Artificial Responsável para Linguística Computacional, Tratamento e Disseminação de Informação (INCT-TILD-IAR). 

\bibliographystyle{source/ACM-Reference-Format}
\bibliography{ref}

\end{document}